%% file: LocalAnalogsPaper_May-6-2021.tex
\documentclass{aastex63}
\submitjournal{ApJ}
\shorttitle{Local Analogs to High-$z$ Galaxies.}
\shortauthors{Moti\~no et al.}

\usepackage{amsmath} 

\usepackage{natbib}
\usepackage[english]{babel} 

\newcommand{\zsun}  {$Z_\odot$}

\def\sun{$_\odot$ }

\newcommand{\akari}    {{\it AKARI}}

\newcommand{\hawc}   {{HAWC+}}

\newcommand{\irac}    {{IRAC}}
\newcommand{\iras}    {{IRAS}}
\newcommand{\irs}      {{IRS}}
\newcommand{\mips}  {{MIPS}}

\newcommand{\pacsi} {{PACS}}
\newcommand{\sof}     {{SOFIA}}
\newcommand{\spire}  {{SPIRE}}
\newcommand{\spitz}  {{\it Spitzer}}
\newcommand{\sdss}  {{\it SDSS}}
\newcommand{\wise}  {{\it WISE}}

\def\iras{{\itshape IRAS\/}}

\def\spitzer{{\itshape Spitzer\/}}

\def\herschel{{\itshape Herschel}}
\def\wise{{\itshape WISE\/}}

\def\ltsima{$\; \buildrel < \over \sim \;$}
\def\simlt{\lower.5ex\hbox{\ltsima}}
\def\gtsima{$\; \buildrel > \over \sim \;$}
\def\simgt{\lower.5ex\hbox{\gtsima}}
\def\kms{\ifmmode{~{\rm km~s^{-1}}}\else{~km s$^{-1}$}\fi}
\def\lsim{\lower0.3em\hbox{$\,\buildrel <\over\sim\,$}}
\def\gsim{\lower0.3em\hbox{$\,\buildrel >\over\sim\,$}}

\def\h2{H$_2$}

\def\arcsec{\mbox{$^{\prime\prime}$}}
\def\arcmin{\mbox{$^\prime$}}

\def\aap{A\&A}
\def\apj{ApJ}
\def\apjl{ApJL}
\def\apjs{ApJS}
\def\aj{AJ}
\def\mnras{MNRAS}
\def\araa{ARA\&A}

\newcommand{\hipe}          {{\sc HIPE}}
\newcommand{\light}          {{\sc Lightning}}
\newcommand{\magphys}  {{\sc MAGPHYS}}

\def\arcsec{\mbox{$^{\prime\prime}$}}
\def\arcmin{\mbox{$^\prime$}}

\begin{document} 
\correspondingauthor{Skarleth Moti\~no Flores}
\email{motioflores@cua.edu}  

  \title{Local Analogs to High-Redshift Galaxies: \\ I. Characterization of Dust Emission and Star Formation History}
  
  \author{Skarleth M. Moti\~no Flores} \affil{Physics Department, The Catholic University of America, 620 Michigan Ave. NE, Washington, DC 20064, USA}
  \author{Tommy Wiklind} \affil{Physics Department, The Catholic University of America, 620 Michigan Ave. NE, Washington, DC 20064}
  \author{Rafael T. Eufrasio} \affil{Department of Physics, University of Arkansas, 248 Physics Building, 825 West Dickson Street, Fayetteville, AR 72701, USA }


%
\begin{abstract} 
Star-forming dwarf galaxies have properties similar to those expected in high-redshift galaxies. Hence, these local galaxies may provide insights into the evolution of the first galaxies, and the physical processes at work. We present a sample of eleven potential local analogs to high-$z$ (LAHz) galaxies. The sample consists of blue compact dwarf galaxies, selected to have spectral energy distributions that fit galaxies at $1.5<z<4$. We use {\sof}-{\hawc} observations combined with optical and near-infrared data to characterize the dust properties, star formation rate (SFR) and star formation histories (SFH) of the sample of LAHz. We employ Bayesian analysis to characterize the dust using two component black-body models. Using the \light\ package we fit the spectral energy distribution of the LAHz galaxies over the FUV-FIR wavelength range, and derive the SFH in five time-steps up to a look-back time of 13.3\,Gyr. Of the eleven LAHz candidates, six galaxies have SFH consistent with no star formation activity at look-back times beyond $1$\,Gyr. The remaining galaxies show residual levels of star formation at ages $\gtrsim$1\,Gyr, making them less suitable as local analogs. The six young galaxies stand out in our sample by having the lowest gas-phase metallicities. They are characterized by warmer dust, having the highest specific SFR, and the highest gas mass fractions. The young age of these six galaxies suggests that merging is less important as a driver of the star formation activity. The six LAHz candidates are promising candidates for studies of the gas dynamics role in driving star formation.
\end{abstract} 



\section{Introduction}\label{intro}

The high-redshift (high-$z$) universe is very different from our local environment: galaxies are physically small with irregular morphologies; they are metal-poor, star formation rates (SFRs) are high and growing exponentially, and the interstellar gas constitutes a major component of the total baryonic mass. However, the faintness and the small sizes make it extremely difficult to study the physical processes in these galaxies in detail. Instead, a significant effort over the last decade has been devoted to identifying low-redshift (low-$z$) analogs to high-$z$ galaxies. (e.g. \citealt{Bian2016}, \citealt{Sebastian2019}).
These analogs to high-$z$ galaxies, are usually local low metallicity, star forming dwarf galaxies. They share many properties of high-redshift galaxies in terms of morphology, physical size, metallicity and star formation (e.g. \citealt{Kaaret2011}, \citealt{Adamo2011}; \citealt{Ostlin2014}). This provides an attractive route to study the properties of high-$z$ galaxies in better detail.

\smallskip  

Several previous studies have defined analogs to high-$z$ galaxies using a variety of approaches. For instance, \cite{Heckman2005}
and \cite{Hoopes2007} found that ultra-violet luminous galaxies (UVLGs) with high surface brightness have characteristics that are remarkably similar to Lyman break galaxies (LBGs)
at high-$z$. UVLGs are rare in the local universe, but a significant sample can be defined when extending the search to $z$$\sim$0.2. These Lyman Break Analogs have $L_{\rm{UV}}\gtrsim 
2 \times 10^{10}$\,L$_{\odot}$, star formation rates (SFR) of 3-30\,M$_{\odot}$\,yr$^{-1}$, sizes of a few kpc, and sub-Solar metallicities. In a similar manner, \cite{Overzier2011} defines 
analogs of high-$z$ LBGs based on their UV luminosity. \cite{Ostlin2014} uses H$\alpha$ equivalent widths as a requirement to identify local galaxies that 
have Ly$\alpha$ luminosities comparable to high redshift Ly${\alpha}$ Emitters (LAEs) and LBGs (the Ly$\alpha$ Reference Sample, or LARS). The galaxies comprising LARS are small, 
metal-poor, $\sim$0.2-0.6\,Z$_{\odot}$, gas-rich, with an average gas mass fraction $\sim$0.4, and SFRs ranging from 0.6 to $\sim$20\,M$_{\odot}$\,yr$^{-1}$. They are morphologically identified as dwarf irregular galaxies (\cite{Pardy2014}). 
Interestingly, most of the LAE and LBG analogs defined by \cite{Ostlin2014} are also far-infrared bright, showing the presence of significant amounts of dust in their interstellar medium, including one galaxy with a Ly$\alpha$ escape fraction of $\sim$12\%.

Another proposed analog for high redshift LAEs is the class of Green Peas (\citealt{Cardamone2009}). 
These are compact, low-mass galaxies, with sub-Solar metallicities. They have SFRs of $\sim$4\,M$_{\odot}$\,yr$^{-1}$, but very high specific star formation rates (\citealt{Izotov2011}). Their properties overlap with Blue Compact Dwarf galaxies (BCDGs) at the high luminosity end of BCDGs, but are otherwise distinct. Green Peas exhibit starburst driven outflows
(\citealt{Yang2017}) and many have a high Ly$\alpha$ escape fraction (\citealt{2021arXiv210408282K}). 
The typical distance of Green Peas is $z\sim$0.2-0.3.

\smallskip

Although these analogs to high-$z$ galaxies are `local' in a cosmological sense, they are still quite distant as far as detailed studies are concerned. The UVLGs and Green Peas are at $z\sim0.2-0.3$, or $D_{\rm{A}} \gtrsim 700$\,Mpc, while the average distance to galaxies in the LARS sample is $\sim$300\,Mpc.
In linear scale, this correspond to $\gtrsim$3.3 and 1.5\,kpc/arcsecond, respectively.
The relatively large distances are necessary because of the low number densities of galaxies fulfilling the criteria for being selected as an LBG and/or LAE analog.

\smallskip

In this paper we present an alternative approach to define local analogs, that allows us to select nearby galaxies at distances $\sim$10-20\,Mpc. This means that scales of 50\,pc can be resolved even with ground-based instruments. Instead of matching properties like luminosity and/or equivalent widths of emission lines, this approach relies on how well the entire UV
to near-infrared spectral energy distribution (SED) of local galaxies match those of high-$z$ galaxies. Using templates of local galaxies spanning a wide variety of morphology,
star formation activity, gas-phase metallicity, as well as AGN activity and environment, we identify those templates that are most successful in fitting high redshift galaxies as potential local analogs to high-$z$ galaxies (LAHz).
The sample of LAHz presented in this work, is selected from 129 local galaxy spectral templates compiled by \cite{Brown2014}. These templates are fit to $\sim$180,000 high redshift
galaxies in the CANDELS (Cosmic Assembly Near-infrared Deep Extragalactic Legacy Survey) fields. It turns out that only a small number of the local templates provide
the best-fit SED to galaxies at $z>2$. All the local templates in this category are designated as blue compact dwarf galaxies. The LAHz are thus scaled-down versions of star forming
high redshift galaxies.

\smallskip

All of the LAHz, as well as most of the galaxies in the Ly$\alpha$ Reference Sample (\citealt{Ostlin2014}) are detected at far-infrared (FIR) wavelengths, suggesting thermal dust
emission. The connection between star formation activity and dust content in galaxies is well-established, and provides information about the evolutionary stage of galaxies. 
In a study involving $\sim$1600 low-redshift galaxies, \cite{DaCunha2010} found a strong correlation between the dust-to-stellar mass ratio and the specific star formation
rate, as well as a strong correlation between the dust-to-SFR ratio and sSFR. These correlations suggests an evolutionary sequence and can be used as diagnostic
for inferring the evolutionary stage of galaxies in combination with other parameters, like the star formation history (SFH).

\smallskip

Observations with the Atacama Large Millimeter Array (ALMA) over the last few years have vastly increased our understanding of the cold and dense interstellar medium of
galaxies in the early universe. Star-forming galaxies on the Main Sequence (MS) at $z$$\sim$3 appear to have dust attenuation properties similar to large and massive
star-forming galaxies in our local universe (\citealt{Fudamoto2017}). Recent results from the {\tt ALPINE} survey (\citealt{LeFevre2020}), however, suggest that the dust
attenuation decreases drastically for galaxies on the MS at $z$$\gtrsim$4 (\citealt{Fudamoto2020}), approaching that of metal-poor dwarf galaxies in the local universe.

\smallskip

The shape of the far-infrared dust SED of high-$z$ galaxies has largely remained unconstrained, as most rest-frame FIR observations at high-$z$ consists of a single photometric
data point. This is also changing thanks to data from ALMA. The FIR SED can now be constructed for galaxies at $z$$\sim$2, where \cite{Pantoni2021} find luminosity-weighted
dust temperatures $\sim$50\,K for 11 galaxies on the MS. \cite{Faisst2020} measured the dust SED for a handful a MS galaxies at $z$$\gtrsim$4,  showing that the peak
temperatures characterizing the dust SEDs are warmer than for star-forming galaxies in the local universe.

\smallskip

In this paper we provide a detailed characterization of the dust in our sample of local analogs to high redshift (LAHz) galaxies. We present new SOFIA HAWC+ observations
which, together with ancillary data, allow us to define the dust SED in detail. This is used to evaluate whether they share properties with galaxies at high redshift, or whether they are similar to local massive star-forming galaxies. 

\smallskip

This paper is organized as follows: In Section.~\ref{Sec:SamSel} we describe the sample definition and selection method. In Section.~\ref{Sec:Data} we describe our \sof  ~observations and ancillary data used. In Section.~\ref{Sec:Modeling} we describe the two modeling procedures used to characterize the dust continuum and the FUV to FIR SED. In Section.~\ref{Results} we present the dust properties and SFHs and in Section.~\ref{Sec:Discussion} discuss the implications of this results. In Section.~\ref{Sec:Sumary} we summarize our results and conclusions about the properties of our sample of LAHz.

\section{Sample Definition}\label{Sec:SamSel}

%
\begin{figure*}[ht!]
\plotone{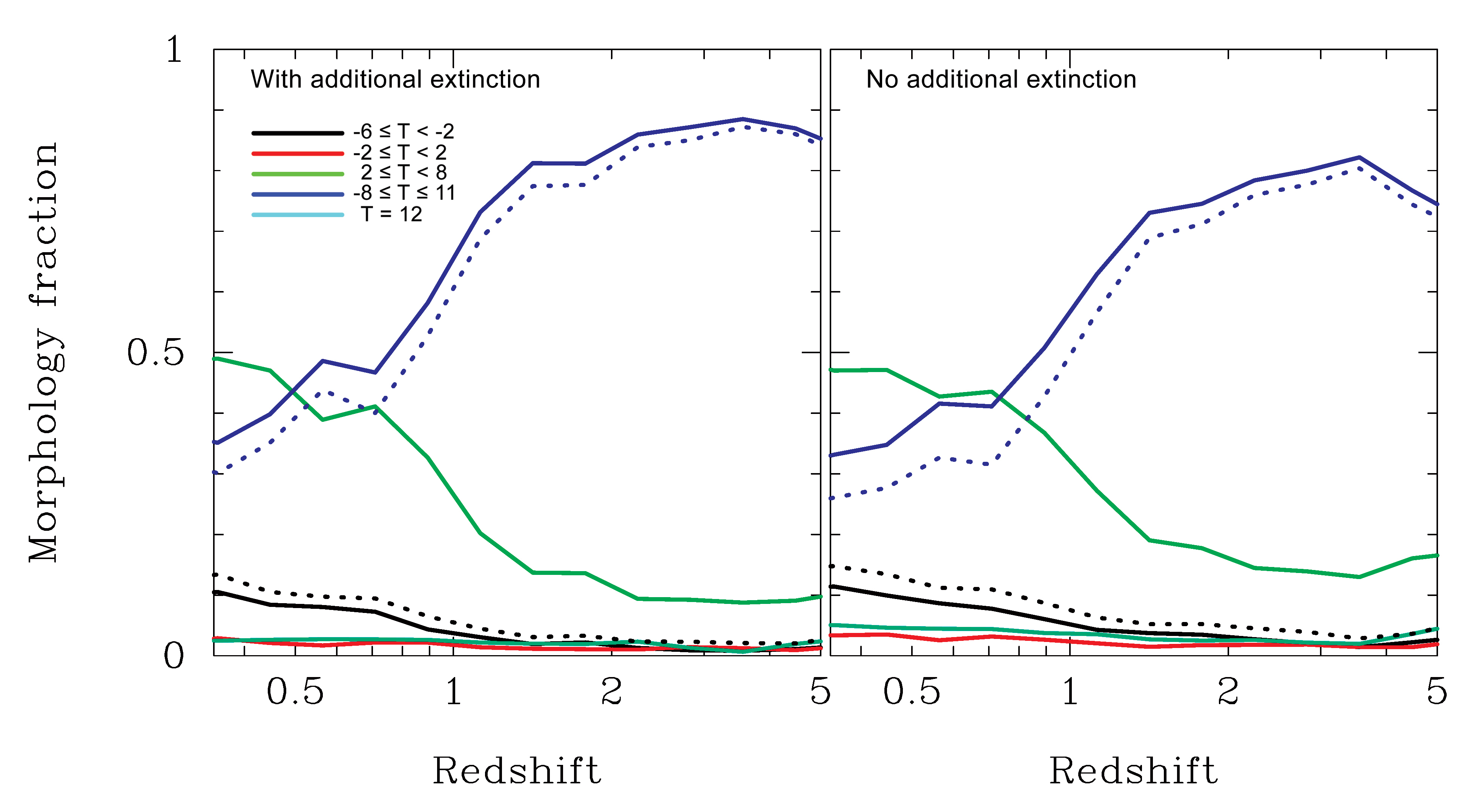}
\caption{The morphology fraction of the \cite{Brown2014} templates, divided into four groups based on morphological T-class. The fraction for each redshift interval represents the best-fit template to CANDELS H-band selected galaxies from all five CANDELS fields. The left panel represents the fractions where we allow additional dust extinction on top of that already implicit in the templates. The right panel we only use the templates, without any additional extinction. The dotted blue line represents the fraction of the eleven templates
selected to be LAHz. The dotted black line is the sum of the two earliest morphological classes, representing quiescent spheroidal galaxies.}\label{fig:morph}
\end{figure*}

\begin{figure}[]
\plotone{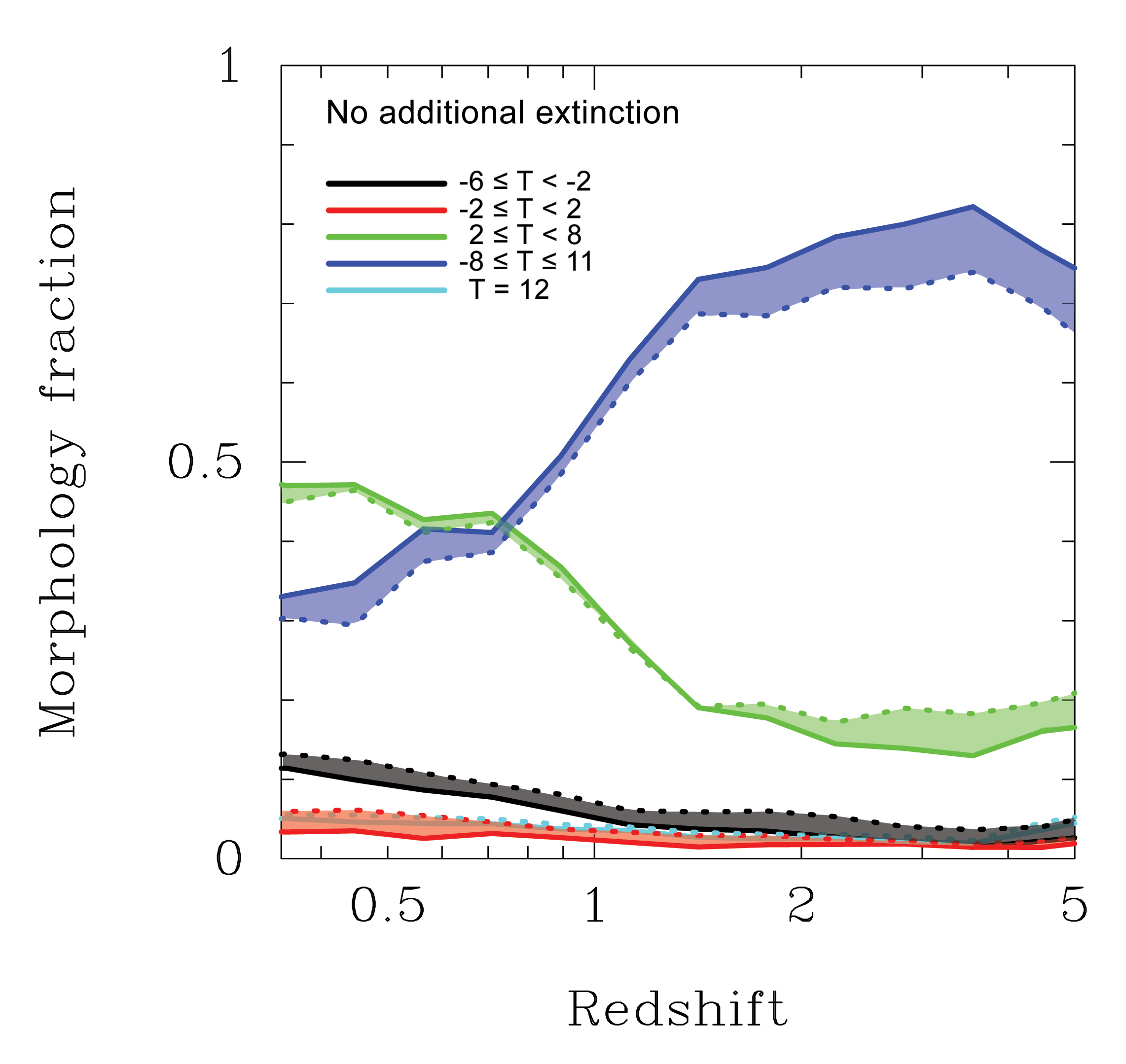}
\caption{Similar to Figure~\ref{fig:morph}. Here the dashed lines correspond to the best-fit SED to CANDELS galaxies when selecting those with $\Delta\chi^2_{\nu} = 3$ relative the
best-fit SEDs shown in Figure~\ref{fig:morph} (shown as full drawn lines). The fraction of irregular galaxies decreases slightly, while the fraction of spiral disk galaxies increases.
The overall dominance of irregular galaxies in the best-fit SEDs is, however, retained.
}\label{fig:chi3}
\end{figure}

\input{Table1_SampleDescription}\label{TableSampleDesc}

\begin{figure*}[!ht] 
\includegraphics [width=\linewidth]{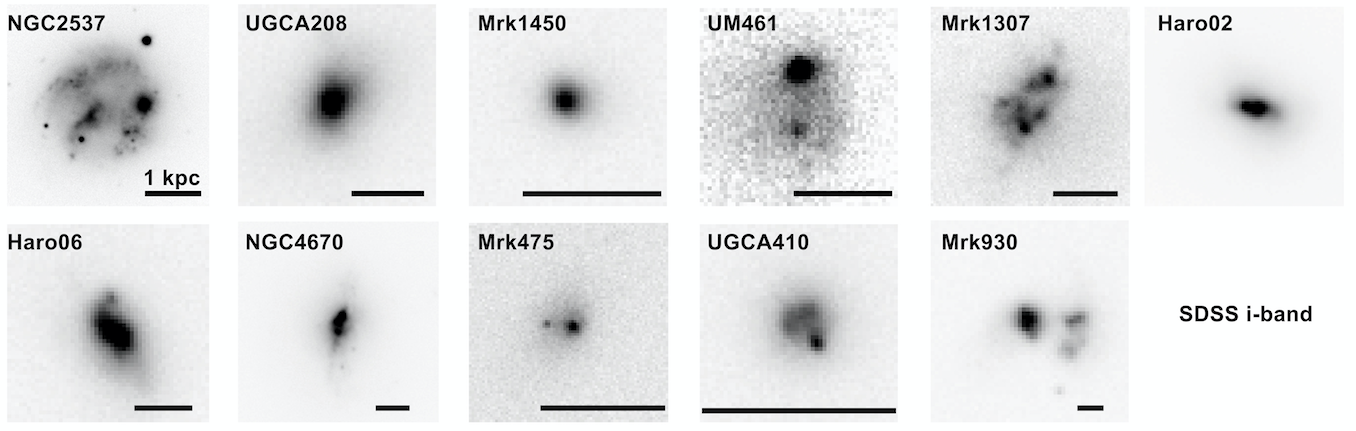}
\caption{SDSS $i$-band images of our sample. All these galaxies are classified as blue compact dwarf galaxies. The bar corresponds to 1 kpc.} \label{sampleBPTD}   
\end{figure*}

Our sample of local analogs to high-$z$ (LAHz) is defined by fitting spectral templates of local galaxies to H-band selected galaxies in the CANDELS survey (\citealt{Grogin2011}; \citealt{Koekemoer2011}), and selecting those templates that provide the best fit to $z\gtrsim2$ galaxies as candidates of being local analogs. 

\smallskip

We use the galaxy templates presented in \citet{Brown2014}. This set of templates contains SEDs covering a wavelength range from UV to the mid-infrared for 129 nearby galaxies.
The templates span a wide range of morphologies and galaxy luminosities; they include ellipticals, lenticulars, spirals, as well as irregular and dwarf galaxies. The sample also contains gravitationally interacting galaxies in various stages of mergers, galaxies with an AGN, and luminous infrared galaxies.
The templates combine ground-based optical spectroscopy with near and mid-infrared spectroscopy from {\spitz}~and {\akari}.
The wavelength coverage is not contiguous and gaps are filled in using \magphys\ (Multi-wavelength Analysis of Galaxy Physical Properties code) modeling (\citealt{DaCunha2008}). The spectral energy distribution from different spectral inputs is normalized and verified with 26 bands of matched-aperture photometry. This allows the mitigation of systematic errors, such as scattered light in \irac\ images and errors in the pre-launch {\wise~W4} filter curve (see \citealt{Brown2014} for details).

\smallskip

For the purpose of establishing which type of templates provide the best-fit SEDs to the high redshift galaxies, we divide the 129 templates into four T-class groups (\citealt{Vaucouleurs1991}) : $-6 \leq T < -2$, $-2 \leq T < 2$, $2 \leq T < 8$, $8 \leq T \leq 11$. These four groups correspond to spheroidal, bulge dominated, star
forming disks, and star forming irregular galaxies. Galaxies without a designated morphology class, are put in T-class 12, peculiar galaxies. The number of templates
in the four main groups are 22, 24, 44, and 20, respectively. In addition, the $T=12$ peculiar group contains 19 templates.

\subsection{High redshift data}{\label{sec:photometry}

The \citet{Brown2014} templates are used to fit to the photometry of high redshift galaxies from the Cosmic Assembly Near-infrared Deep Extragalactic Legacy Survey (CANDELS). 
CANDELS is a treasury program on Hubble Space Telescope (HST), and provide deep multi-wavelength imaging data in five legacy deep fields: GOODS-South, GOODS-North, UDS, COSMOS and EGS (\citealt{Grogin2011}; \citealt{Koekemoer2011}). In this study we use H-band selected catalogs from all five fields. Details on the data and photometry at different wavelength bands are given for; GOODS-S (\citealt{Guo2013}; \citealt{Santini2015}), GOODS-N (\citealt{Barro2019}), UDS (\citealt{Galametz2013} ; \citealt{Santini2015}), COSMOS (\citealt{Nayyeri2017}), and EGS (\citealt{Stefanon2017}). 

\smallskip

The CANDELS catalogs contain photometric data from the UV to near- and mid-infrared wavelengths, observed in broad, medium, and in some cases, narrow band filters. The catalogs also contain derived physical parameters, such as photometric redshift, stellar mass and star formation rates, compiled using several different teams applying different SED methods, star formation histories and different templates (e.g. \citealt{Dahlen2013}; \citealt{Mobasher2015}; \citealt{Santini2015}). The optical (HST/ACS) and near-IR (HST/WFC3) data are consistently combined with the mid-IR data (\spitz/\irac) and ground-based observations (UV and K-band). All of the photometric catalogs were selected in the HST/WFC3 F160W band using SExtractor (\citealt{Bertin1996}).
The catalogs for the five different CANDELS fields contain a total of 185,956 galaxies. The number of photometric bands range from 18 (GOODS-S) to 42 (CANDELS COSMOS). The latter catalog includes extensive narrow- and medium-band photometric bands. The mid-infrared data are from {\spitz/\irac} (\citealt{Ashby2013}).

\subsection{SED fitting}

All 129 of the \citet{Brown2014} templates are used in the SED fitting of the 185,956 CANDELS galaxies. The fits are done in two different modes: keeping the extinction equal to the intrinsic dust extinction of the templates, and allowing additional extinction on top of the intrinsic.  The additional extinction is parametrized through the $E(B-V)$ color index. When treated as a free parameter, we allowed additional dust extinction up to $E(B-V) = 0.6$. We use the \cite{Calzetti2000}  attenuation law for the additional extinction. An initial test was done keeping the photometric redshift as a free parameter. The redshifts from the SED fits were similar to those presented in the CANDELS catalogs. For the reminder of the analysis, we keep the redshift fixed at the CANDELS values, which includes spectroscopic redshifts whenever available.

\smallskip

The SED fitting procedure involved redshifting the template to the redshift of each galaxy, then each galaxy template is convolved with the filter response functions appropriate for a given CANDELS data set. The template is then fitted to the observed flux densities using $\chi^2$ minimization. If additional extinction is included, we repeat this process for $E(B-V)$ values ranging from 0.0 to 0.6, in steps of 0.01.

\smallskip

We divide the CANDELS galaxies into 13 redshift bins between $0.3 \leq z \leq 5$ ($\Delta z = 0.36$). The total number of galaxies in each redshift bin range from $\sim$2,000 at z$\sim0.3$ to $\sim$10,000 at $z\sim2$. For each redshift bin we derive the number of best-fit SEDs for each morphological group as defined above. In Fig.~\ref{fig:morph} we plot the fraction of best-fit SEDs for the four morphological classes as a function of redshift. From the figure it is clear that the Irregular class ($6 \leq T \leq 11$), provides about 35\% of the best-fit templates at $z\sim$0.3, and increase to dominate at $z\gtrsim$1. At $z\gtrsim$2, these templates account for 85-90\% of all best-fit templates. Of the 20 templates in the $6 \leq T \leq 11$ morphological class, 11 of them provide the most frequent best-fit SED (dotted line in Fig.~\ref{fig:morph}). These 11 galaxies are selected as our sample of potential Local Analog candidates.

\smallskip

In order to test the robustness of the morphological fractions shown in Fig.~\ref{fig:morph}, we selected the template galaxy with $\Delta\chi_{\nu}^2 = 3$ relative to the best-fit SED. The resulting morphological fraction as a function of redshift is shown in Fig.~\ref{fig:chi3}. At $z\gtrsim1.5$, the fraction of the $6 \leq T \leq 11$ morphological class decreases slightly, while the fraction of star forming spiral disks increases. However, the overall dominance of irregular galaxies as the best templates for high redshift galaxies remains. This shows that the result is robust and local templates with T-class morphologies corresponding to irregular star forming galaxies are best at fitting the SED of high redshift galaxies at $z\gtrsim1$.

\smallskip

This result is not surprising as galaxies in the $6 \leq T \leq 11$ morphological class share many properties with star forming galaxies at high redshift (e.g. \citealt{Elmegreen2009}; \citealt{Elmegreen2012}). The general properties of the 11 candidates are summarized in Table~\ref{SampleDescription}; they have small physical sizes ($1-2$\,kpc), they are actively forming stars, and they typically have a high gas mass fraction. Their gas-phase metallicity is low, with an average metallicity of $12 + \log{O/H} = 8.11 \pm 0.20$, or $\sim$0.26\,Z$_{\odot}$. In Fig.~\ref{sampleBPTD} we show the Sloan Digital Sky Survey (SDSS) i-band images of the 11 candidates. From these images it is clear that they have irregular morphology with prominent star forming clumps.

\smallskip

All 11 galaxies fall under the classification of Blue Compact Dwarf Galaxies (BCDGs; e.g. \citealt{Sargent1970}; \citealt{Kunth1988}; \citealt{Thuan2005}). BCDGs have previously been proposed to be analogs of high redshift galaxies (e.g. \citealt{Hoopes2007}). As a group they share many properties, such as small sizes, clumpy and intense star formation. They show strong optical emission lines, often exhibiting high excitation (\citealt{Thuan2005}). There are, however, differences among the BCDGs; some show the star forming regions superposed on faint extended stellar emission, suggesting star formation extended over longer time periods (e.g. \citealt{Noeske2003}), while others seems to be truly young systems (e.g. \citealt{Hunt2014}, \citealt{Papaderos2008}). The gas-phase metallicities can range from extremely low, like SBS\,0335-052E and IZw18, with $Z\approx0.03$ \cite{Izotov2005}, to $Z\approx$0.3\,Z$_{\odot}$.
One feature that most BCDGs share is the presence of dust. Even the low-metallicity systems SBS\,0335-052E and IZw18 show thermal dust emission, even when detectable CO emission is absent (\citealt{Hunt2014}). In the reminder of this paper we will characterize the dust emission and the star formation history of the 11 Local Analog candidates defined above, to see whether they can be viewed as scaled down versions of high redshift galaxies.

%
\begin{figure*}[] 
\includegraphics [width=\linewidth]{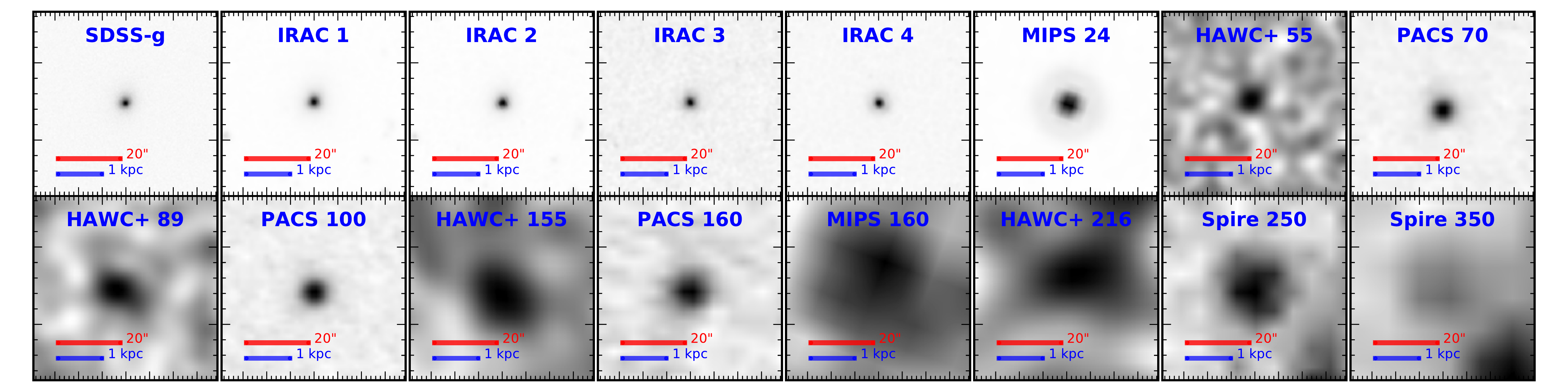} 
\caption{Optical to far-infrared images of Mrk 1450. The red bar corresponds to $20^{\prime\prime}$, and the blue bar indicates 1 kpc.}
\label{MRK1450}    
\end{figure*}


\begin{table*} [!htb]
\centering
\caption{Multi-wavelength data set\tablenotemark{1}}
\label{SampleDataA}
\input{Table2_DataAvailable.tex}
\tablenotetext{1}{W stands for \wise, \irs~is for Infrared Spectrograph of \spitzer, M for \mips, S for {\sof}, H for \herschel\ and I for \iras, A for \akari.} 
\end{table*}

\section{Data}\label{Sec:Data}
In this section we describe the data reduction process we used to produce maps of our target galaxies, and the methods to obtain photometric estimates. 

\smallskip

In order to characterize the 11 nearby galaxies that make up our sample of potential local analogs, we compile broad-band photometric data, covering UV to FIR. We supplement
the existing UV-NIR photometric data from \cite{Brown2014} with FIR observations of five galaxies in our sample using the Stratospheric Observatory for Infrared Astronomy ({\sof}).
We combined these with archived infrared data from \herschel, \spitz, \akari, and \wise.
A listing of the infrared data sets used in our analysis, indicating the different telescope and instrument combinations, is given in Table~\ref{SampleDataA}. In
Figure~\ref{MRK1450} we show an example of imaging of one of our galaxies, Mrk\,1450. The images shown consist of our {\sof}-\hawc\ observations, along with archival data from \spitz~and \sdss~band-g (Sloan Digital Sky Survey \citealt{Gunn2006}) and \herschel\ \pacsi~and \spire~instruments. Mrk\,1450 is well-detected in all bands, but starts to fade in the \spire~$350\,\mu$m band, and is undetected in the \spire~$500\,\,u$m band.

\subsection{\sof~Observations}

The Stratospheric Observatory for Infrared Astronomy ({\sof}: \citealt{Young2012}, \citealt{Temi2014} ) is a partnership between NASA and the German Aerospace Center (DLR), consisting of an extensively modified Boeing 747SP aircraft carrying a 2.7-meter (106 inch) reflecting telescope (with an effective diameter of 2.5 meters or 100 inches). We were
granted $\sim$8 hours of observing time with {\sof}; on Cycle 6 under the program 06-0222 (PI: T. Wiklind). We observed 5 galaxies of our sample, using the High-resolution Airborne Wideband Camera-plus instrument (\hawc;\  \citealt{Vaillancourt2007}; \citealt{Dowell2010}; \citealt{Harper2018}), covering all available filters to derive the far infrared spectral energy distribution. 

\smallskip

The \hawc\ instrument has 5 bands identified as A, B, C, D and E. We observed in bands A, C, D, and E at $53$, $89$, $154$ and $214\,\mu m$, which have beam sizes (full width at half maximum, FWHM) of $4.85\arcsec$, $7.80\arcsec$, $13.6\arcsec$ and $18.2\arcsec$ respectively.  Band B, was not available during Cycle 6. 
The galaxies Mrk 1450, NGC 4670, and Haro 06  were observed in all 4 bands available. Due to time constraints, Mrk 1307 and NGC 2537 were only observed at $53$, $89$, and $155 \,\mu m$ bands. The data was reduced to obtain the total intensity maps using the \hawc\ Data Reduction Pipeline v1.3.0 as detailed in \cite{Harper2018}.

\begin{table*} [!htb]
\centering
\caption{UV - FIR Photometric Data\tablenotemark{1}}
\label{PhotometryTable}  
\input{Table3_PhotSOFIA.tex} \label{TablePhotometry}
\tablenotetext{1}{This is an abbreviated version of the complete table of flux values used for the analysis. The complete table is available in machine readable form in the digital version of this paper. The full table contains 49 columns, indicating the telescope and instrument/band for each measurement and the uncertainty. Fluxes are given in mJy.}
\end{table*}

\subsubsection{\sof/\hawc Photometry} \label{drSOFIA}

The \hawc\ data is presented in 5 levels: Level 0 is raw data, Level 1 is the demodulated (chop not subtracted) data, after instrument corrections (flat-fielded, dark-subtracted, bad pixel removed) becomes Level 2, level 3 is obtained after flux calibration and telluric correction, and the combined observations from level 3 produces the level 4 data. More details on the pipeline process can be found in the data handbooks (\citealt{Gordon2018}).

\smallskip

We started our analysis with level 4 maps in units of Jy/pix, and performed annular photometry using the \hipe~environment. A circular aperture, centered on the source emission, gives
the flux + background emission; an annular region, free of emission and centered on the inner aperture, provides an estimate of the background.
To determine the photometric errors we placed six circular apertures of the same radius as the inner source aperture, at equidistant, flux-free, areas around the source. The average flux of the six apertures ($Bg_{ave}$) is then subtracted from the source flux:

\begin{equation}
 F_{source} = (f_{tot} ) - Bg_{ave}.
\end{equation}

As a further test, we estimated the integrated flux using {\sc Casa}\footnote{\href{https://casa.nrao.edu/}{https://casa.nrao.edu/}}, and the built-in two-dimensional gaussian procedure.
This produced integrated fluxes very similar to the ones obtained using aperture photometry.

\smallskip

To determine the photometry errors we used 6 photometry apertures around the source, taking care to avoid any other source in the background, then then uncertainty of the measurement was defined as the standard deviation of the background apertures. The total error is computed as the addition on quadrature of the uncertainty of the measurement $Bg_{std}$ plus the absolute calibration error, the absolute calibration percentages are listed in Table~\ref{TotAbsError}.

\begin{equation}
 F_{unc} = \sqrt{(F_{source}*Abs_{cal})^2 + {Bg_{std}}^{2}}.
\end{equation}

The \sof/\hawc~and \pacs~photometry measure and uncertainties is shown in Table~\ref{TablePhotometry} (the full table is available in machine readable form in the 
digital version of this paper).


\begin{table} [] 
\centering
\caption{Fractional calibration error.} \label{TotAbsError}
\begin{tabular}{l c c}
        \hline   \hline   
        Telescope        &  Instrument   &  Abs. Calibration     \\
        \hline
               \sof           &  \hawc          &    $15\%$                 \\
               \herschel\       &  \pacsi          &     $10\%$              \\
               \herschel\       &  \spire           &     $15\%$              \\
               \wise        &   W1-W4       &     $6\%$                       \\
               \akari       &   N:60,160,WIDE:S,L  &     $10\%$     \\
               \spitz       &   \irs               &     $5\%$                       \\
               \spitz       &   \irac             &     $5\%$                       \\
               \spitz       &   \mips           &     $5\%$                        \\
             \hline  \hline
\end{tabular}
\end{table}

\subsection{Ancillary UV to MIR data}

The \cite{Brown2014} templates contain SEDs covering a wavelength range from UV to the mid-infrared for 129 nearby galaxies.
As described in section~\ref{Sec:SamSel}, the SEDs of these templates have been normalized and verified with matched-aperture photometry.
They provide accurate photometric data, and since our LAHz sample is drawn from these templates, we use the photometric data as presented in \cite{Brown2014}.

\subsection{Ancillary FIR Data}
Some galaxies in our sample were observed  with the {\herschel\  Space Observatory},
using \pacsi\ at $70$, $100$, and $160\,\mu m$, with a FOV of  $1.75\arcmin \times 3.5\arcmin$, and \spire~at $250$, $350$, $500\,\mu m$ with a FOV $4\arcmin \times 8\arcmin$. 

\smallskip

The \herschel\ Dwarf Galaxy survey (\citealt{Madden2013}; \citealt{Ruyer2013}) provide photometric and spectroscopic FIR data on 50 dwarf galaxies.
This survey includes data for some of the galaxies in our survey: Mrk~930, Mrk~1450 and UM~461, which were observed with all bands of both \pacsi~and \spire~instruments. 
A limited set of \herschel~photometric data is available  for UGCA~410 (also knowns as  Mrk~140) and Haro~02 (\citealt{Herschel2013}). See the full Table~\ref{PhotometryTable}
for a complete listing of the FIR photometric data for our sample.
For consistency in our flux estimates, we reduced the available \herschel\ \pacsi\ and \spire\ data using the \herschel\ pipeline for total intensity photometric measurements.
We used the {\herschel\ Interactive Processing Environment} \hipe~(\citealt{Ott2010}) with version 7 of the photometric calibration.

\subsubsection{\herschel\  Photometry}\label{drHerschel}

We use level 2.5 or 3.0 maps for our photometry estimates. These maps are created from a combination of different observations of the same target and have improved S/N levels.

For more detail about \herschel~products see \pacsi~Data Reduction Guide\footnote{\href{http://herschel.esac.esa.int/hcss-doc-15.0/print/pacs_phot/pacs_phot.pdf\#pacs_phot}{http://herschel.esac.esa.int/hcss/pacsDRG}} .

\smallskip

To determine the total intensity in \herschel\ \pacsi~$70$, $100$, and $160\,\mu$m maps, we use the \pacsi~- Photometry Script,  which is part of \herschel~common science system (HCSS) provided in the \hipe\ environment. We started with the maps: Photometry Color High Pass Filter Map  (HPPHPFMAPR) level 2.5 or 3.0, depending of availability. We modified parameters in the algorithm in orden to use the optimal aperture radius. The final outputs of the algorithm, are the peak flux in Jy/pix, the flux measured by the aperture photometry in Jy, and the aperture corrected flux in Jy.  \spire~photometry maps can be in two different observation mode: {\it point source} (psrc) in units of Jy/beam or extended source (extsrc) in units of MJy/sr, and are identified as PSW, PMW, PLW that stands for Photometry-Short/Medium/Long-Wavelength, corresponding to 250, 350 and $500\,\mu$m respectively.  Here we used the psrcPSW, psrcPMW and psrcPLW level 2.5 maps. 

\smallskip

Similarly to \pacsi, for  \spire~photometry maps, we use the \spire~Photometry Point Source Script provided within \hipe. This, uses the sourceExtractorSussextractor task to identify the source, and applies the Color Correction for point sources. The \herschel~pipeline assumes $\alpha=\,-1$, we assumed $\alpha\,=\,2.0$.
 
%
\section{Methodology}\label{Sec:Modeling}

In this section we describe our modeling approach to characterize the dust continuum and the stellar emission. We use two different methods; modified black-body, and the \light\ fitting package (\citealt{Eufrasio2017}).
Both approaches allow for a determination of the far-infrared luminosity and dust mass for each galaxy. The modified black body models are used to also derive luminosity-weighted, dust temperatures, while the
\light\ method allows a determination of the star formation history (Section~\ref{sec:light}).

\subsection{Black-body Modeling}\label{bbmodel}

The simplest characterization of dust emission is to fit it with a black-body $F_{\nu} \propto B_{\nu}(T)$. In a more realistic situation, this is modified by taking the transfer of radiation into account.
This can be done by using a so-called modified black-body curve, $F_{\nu} \propto (1 - \exp(-\tau)) B_{\nu}(T)$.
We will use the expression:
 \begin{equation}
    mBB (T)=   \Omega  B_{\nu}(T)  \left(1 - \exp{\left[-\left(\frac{\lambda_{0}}{\lambda}\right)^{\beta}\right]}\right),
\end{equation}
where $B_{\nu}(T)$ is the Planck function, and $\Omega$ the solid angle extended by the emission. The opacity is parametrized as $\tau = (\lambda_0/\lambda)^{\beta}$, where $\lambda_0$ is the wavelength where the optical depth equals unity (e.g. \citealt{Draine2006}), and the coefficient $\beta$ represents the opacity spectral  index (\citealt{Kovacs2010}). Observationally, $\beta$ has been derived to be between 1 and 2 (cf. \citealt{Hildebrand1983}; \citealt{Casey2012}). Data from the Planck mission provides a robust derivation of $\beta$ in the Milky Way, leading to $\beta = 1.8 \pm 0.1$ (\citealt{2011A&A...536A..21P}). For extragalactic objects it is common to use $\beta$ in the range $1.3-2.0$.
Here we will adopt the commonly used value $\beta = 1.5$, and keep it constant in our fits. We did try fitting with different $\beta$ values, but did not find that it changed our results in any significant way. We also kept the wavelength where the optical depth is unity constant at $\lambda_0 = 100\mu$m. The two parameters $\beta$ and $\lambda_0$ affects the flux on the long wavelength Rayleigh-Jeans part of the dust SED relative to the short wavelength part; for a fixed dust temperature, an increasing $\lambda_0$ leads to a relative increase in the Rayleigh-Jeans part, while an increase in $\beta$ leads to a decrease of the Rayleigh-Jeans part relative the Wien part of the SED.

When fitting a modified black body curve to flux measurements at $\lambda \gtrsim 60\mu$m, most galaxies exhibit a mid-infrared excess at wavelengths shorter than 60$\mu$m (e.g. \citealt{Casey2012}). This excess can be attributed to the presence of a warmer dust component than the one characterizing the longer wavelengths, or an additional heating source, such as an AGN (cf. \citealt{ScovilleKwan1976}). 
The observed mid-IR emission can be modeled as a power law, with a cut-off wavelength $\lambda_c$ (cf. \citealt{Casey2012}), or as second, dust component characterized by a warmer dust temperature. 
Since we do not expect a significant AGN activity in the dwarf galaxies making up our sample, we will adopt the latter case, and use two modified black body curves, with the same $\beta$ and $\lambda_0$, but with different dust temperatures.
Our dust SED to be fitted to the observed data is then:

\begin{equation}
\begin{split}
    mBB=   \Omega_1 B_{\nu}(T_{1})  \left(1 - \exp{\left[-\left(\frac{\lambda_{0}}{\lambda}\right)^{\beta}\right]}\right)   \\
+  \Omega_2 B_{\nu}(T_{2})   \left(1 - \exp{\left[-\left(\frac{\lambda_{0}}{\lambda}\right)^{\beta}\right]}\right).
\end{split}
\end{equation}

With the opacity spectral index kept fixed at $\beta = 1.5$, and the critical wavelength at $\lambda_0 = 100\mu$m, the modified black body curve contains four free parameters: two dust temperatures $T_1$ and $T_2$, and two normalization constants containing the solid angle of the source $\Omega$. We use the {\it Levenberg-Marquardt} fitting algorithm. We explored the fitting using values for the dust opacity coefficient $\beta$, ranging from 1.2\,$-$\,2.0, and for $\lambda_0$ in the range $50-200\mu$m. We found no significant variation in the derived dust temperatures, and we conclude that a fixed $\beta =1.5$ with $\lambda_0 = 100\mu$m is sufficient and the models will be characterized by the temperatures and
normalization coefficients. 

\smallskip
 
We employ Bayesian statistics inference in multi-parameter spaces to determine the most probable set of model parameters $\phi$, of a model $mBB_{i}$ in comparison to the data $D_{i}$. Bayesian inference is based the Bayes' Theorem, which establish the conditional probability of the parameters $\phi$ given the data $D_{i}$ as:
\begin{equation}
 p( \phi | D_{i}) = \frac{ p(D_{i} | \phi ) * p(\phi  ) } { p(D_{i}  )},
\end{equation}
where the likelihood of the model is $p(D_{i} | \phi )$. Then $p( \phi | D_{i} ) $ is the posterior probability, $ p(\phi )$ is the priors which is the knowledge we have about the parameters or problem. $p(D_{i} )$ is used to normalize the posterior probability.
The likelihood is assumed to be: 
\begin{equation*}
p \propto  \exp{\left ( - \chi^2 /2 \right)}.
\end{equation*}
By maximizing the posterior probability we will determine each $\phi$ parameters best values for the modified black-body models $mBB_{i}$. Since we are assuming a fixed $\beta=1.5$ and $\lambda_0 = 100\mu$m for both components,
the  $\phi$ parameters are the temperatures $T_1$, $T_2$, and normalization constants $\Omega_{1}$ and $\Omega_{2}$.
The priors are defined by allowing the $\phi$ parameters to vary through a very wide range initially, to get a posterior probability and then bayesian analysis allows us to update the knowledge of the probability distribution of the parameters and apply new priors multiplying by flat priors that reduce the range of the parameters space. This allows us to reduce the region covered by the parameter and obtain a better resolution of the parameter grid.
Inspection of the marginalized probability distribution of each parameter-space, allows us to determine the appropriate range in which the best set of solutions are located, as well as the most likely value. We can also define the confidence interval
for each parameter.

\subsubsection{Dust masses}

The dust mass can be derived from the flux density of optically thin emission.
\begin{equation}
M_{\rm dust}  =  \frac{S_{\nu}\,D^2}{\kappa_{\nu}\,B_{\nu}(T_d)},
\end{equation}
where, $S_{\nu}$ is the observed flux density, $B_{\nu}(T_{\rm d})$ the black-body emission at a given temperature, $D$ is the distance, and $\kappa_{\nu}$ is the dust absorption coefficient for dust grains at frequency $\nu$.
The dust absorption coefficient is a parameter encompassing both the dust grain composition and the grain size distribution, and is difficult to derive observationally (e.g. \citealt{Gordon2014}). \cite{WeingartnerDraine2001} derived
$\kappa_{\nu}$ for different $R_V$ values for Milky Way dust. We will use the average $\kappa_{\nu}$ for $R_V = 3.1, 4.0$ and $5.5$ at $\lambda = 250\mu$m. Using updated values, we have $\kappa_{250} = 0.40$\,m$^2$\,kg$^{-1}$. We do not have observed flux values at $\lambda = 250\mu$m for all of our galaxies. Instead we use the flux density provided by our mBB fits to the overall dust SED.

Hence, the dust mass, using the flux density at 250$\mu$m, can be expressed as: 

\begin{equation}
\begin{split}
\frac{M_{\rm dust}}{M_{\odot}} = 188.4\,\left(\frac{S_{\nu}}{\rm {Jy}}\right)\,\left(\frac{D}{\rm {Mpc}}\right)^2\,\frac{\left(\exp{\left(57.55/ T\right)} - 1\right)}{\kappa_{250}}.
\end{split}
\end{equation}
The dust masses derived in this way are listed in Table~\ref{DustTemperatures} and discussed in Section~\ref{LIRandMasses}.

\subsection{ \light\ SED Fits}\label{sec:light}

We performed SED fitting using \light\ (\citealt{Eufrasio2017}, ASCL:1711.009\nocite{2017ascl.soft11009E}\footnote{\href{https://github.com/rafaeleufrasio/lightning}{https://github.com/rafaeleufrasio/lightning}}), which fits non-parametric star formation histories (SFHs) in discrete, fixed-width, stellar age bins. We made use of the most recent update to \light, which uses an adaptive Markov Chain Monte Carlo (MCMC) algorithm (Doore et al. 2021, in prep). For this work we have used SFH bins of 0$-$10~Myr, 10$-$100~Myr, 100~Myr$-$1~Gyr, 1$-$5~Gyr, 5$-$13.3~Gyr, assumed a Kroupa IMF (\citealt{Kroupa2001}) and a metallicity of 0.2\zsun. We used PEGASE (\citealt{FiocRocca1997A}; \citealt{1999astro.ph.12179F}) stellar populations. For intrinsic attenuation, we adopted a \cite{Calzetti2000} extinction law, modified as in \cite{Noll2009} to include a UV bump at 2175~\AA\ and a parameter $\delta$ to control the UV slope. \cite{DraineLi2007} Milky Way dust emission was used. For a more thorough description of the prescriptions adopted by \light, we refer the reader to \cite{Eufrasio2017}.

\smallskip

The IR photometry previously described described was combined with literature SEDs from \cite{Brown2014}, producing SED with 45 data-points from the FUV to the FIR (including {\it GALEX}, {\it Swift} UVOT, SDSS, 2MASS, \wise, \spitzer, \sof, \iras, \akari, and \herschel). Our fits explore a 13-dimensional parameter space with 5 parameters regarding the SFH intensities, $\psi_i\ge0$, 3 parameters ($0\le\tau_V^{\rm DIFF}\le3$, $-2.3\le\delta\le0.4$, and $0\le\tau_V^{\rm BC}\le4$) controlling the dust extinction, and 5 dust emission parameters ($0\le\alpha\le4$, $0.1\le U_{\rm min}\le25$, $10^3\le U_{\rm max}\le3\times10^5$, $0\le\gamma\le1$, and $0.0047\le q_{\rm PAH}\le0.0458$. We run the adaptive MCMC algorithm for $3\times10^5$ steps and derive statistics from the last 5,000 steps of the chain. All the derived quantities quoted here (SFHs, SFRs, M$_{\star}$, M$_{dust}$) are derived from these last 5,000 steps.

%
\section{Results} \label{Results}

We derive properties of the dust in our LAHz galaxies using two different methods. Some of the derived parameters are duplicated in the two methods, such as the FIR luminosity and
dust mass. Others are unique for each method. For instance, the modified black body models provide luminosity-weighted dust temperatures, which the \light\ models do not. The \light\ models
provide an estimate of the SFR integrated over the last 100\,Myr, and a parametrized star formation history (SFH).
The derived parameters from the modified back body models are presented in Tabel~\ref{DustTemperatures}, and those from the \light\ models in Table~\ref{LIGHTNINGTable}.
\begin{figure*}[!ht] 
\begin{center}
\title{Bayesian Estimation of Black-body Models Best Fits}
\includegraphics [width=\linewidth]{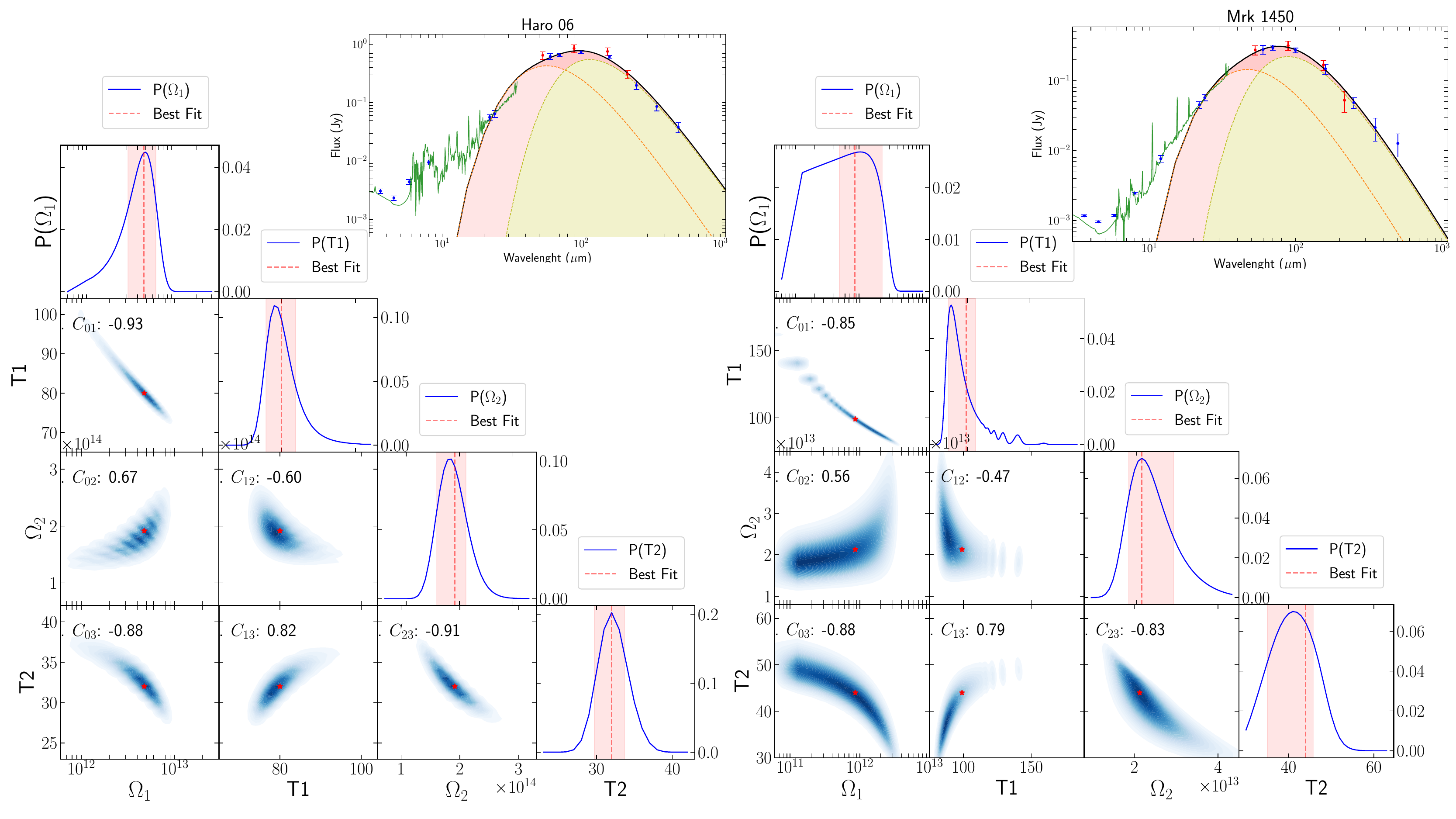}
\caption{Corner plots for the modified black-body models of the SED of Haro~06 and Mrk~1450. In each panel, in the bottom: The 2D scatter plots (blue area) of the values recovered from the four parameters grid are shown for the two-by-two combination of parameters (two temperatures and two normalization constants). On top are the histograms for all 4 parameters we display Bayesian 1D marginalized likelihoods. The best-fit values to the observed data in the histograms are shown in red and the light-red area encloses 68 $\%$ of the likelihood. Top right: Best fit model in black, shows the two dust components, cold component in yellow and warm in light-red. The red are the \sof~data points, and blue data points are the ancillary data available. The green solid line is the \cite{Brown2014} spectral data, which is not part of the fits, however is in good agreement with the modified black-body model. }
\label{BBFitsDiagnosticPlots2}   
\end{center}
\end{figure*}
\subsection{Modified Black-Body models}

The black-body modeling described in section \ref{bbmodel} show that a reasonable description of the dust SED can be achieved by using a two-component modified black-body fit.
The warm component is less well constrained than the cold component due to a smaller number of photometric data points at wavelengths $\lambda \lesssim 60\mu$m. Nevertheless, the
model SED joins smoothly with the near- and mid-IR fluxes of the \cite{Brown2014} templates.

\smallskip

In Figure~\ref{BBFitsDiagnosticPlots2} we show multi-parameter-space diagnostic plots for two galaxies in our sample: Haro~06 and Mrk~1450. We can see the marginalized distribution for each $\phi$ parameter in the upper frame of each column, the lower frames show the marginalized probability in blue and the correlation values between each two parameters. The value that maximizes the $\chi^2 $ distribution (likelihood) for each parameter is indicated with a red star; this is also shown as a red dashed line in the upper frames for each parameter. The confidence interval is shown in light red and indicates the 16 and 84 percentiles in the probability distribution, which imply that the $68\%$ of the probability is inside that region.

\smallskip

This visualization of the probability distributions allow us to understand how the different parameter relate. For instance we can see how the temperature $T_1$ and the constant  $\Omega_{1}$ are anti-correlated, while the two constants $\Omega_{1}$ and $\Omega_{2}$ are positively correlated for both galaxies. 
In the upper right box we see the best fit model indicating the two black-body components representing a warm and a cold dust component. The corresponding \cite{Brown2014} template is shown as a reference, but it is not part of the fit. The resulting best-fit model agrees smoothly with the Brown templates.
%
\begin{figure*}[htb] 
\begin{center}
\includegraphics [width=\linewidth]{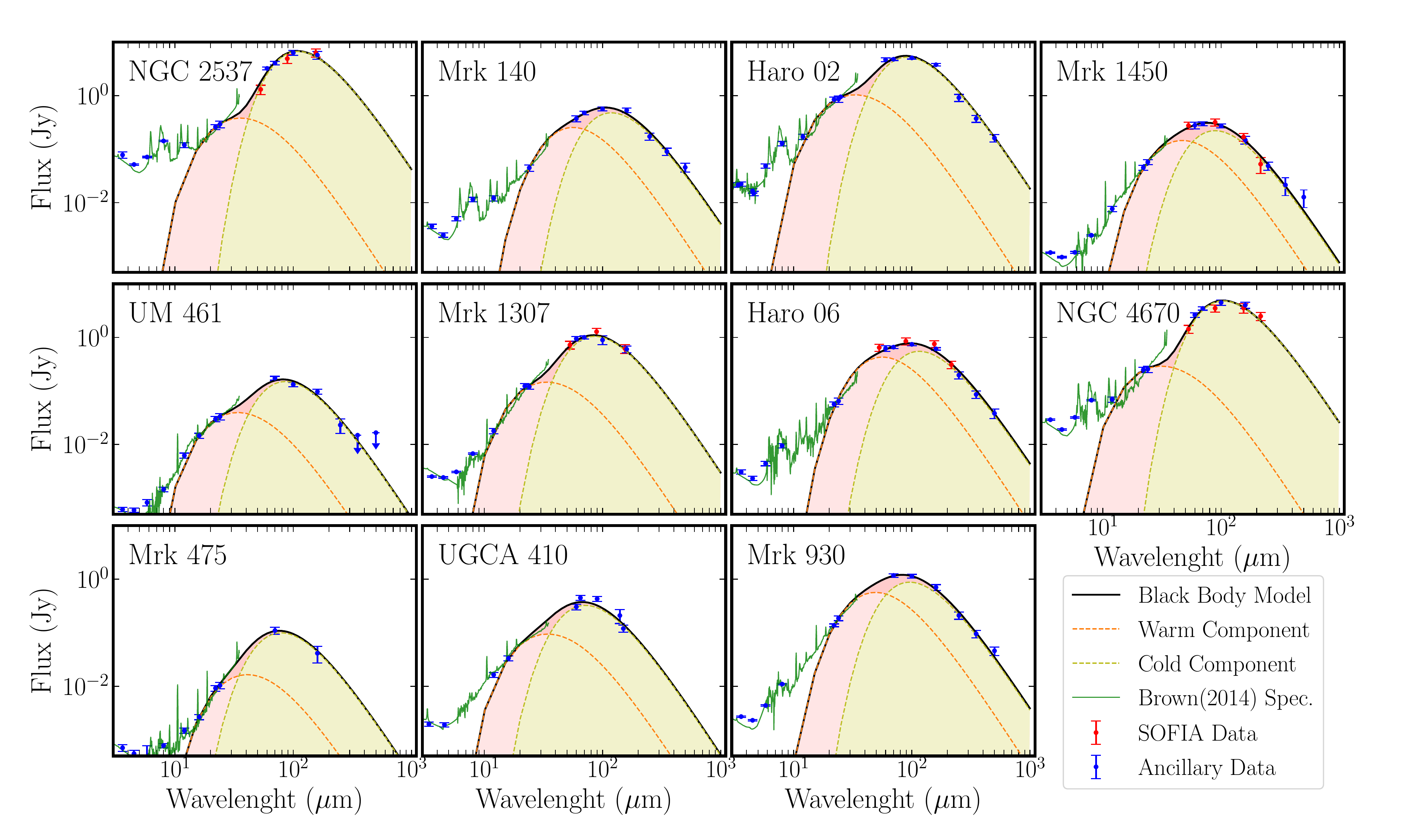}
\caption{ Black-body best fits: LAHz dust emission. The best modified black-body fit shown in black solid line, the light-green and light-red dash lines are the warm and cold components. \sof~(53, 89, 155, $216\,\mu$m) data points in red, ancillary data points in blue \herschel~(60, 100, 160, 250, 350, $500\,\mu$m) data points and \spitz-\mips~(24, 60, 100, 160 $\mu$m) in blue where used when \sof~or \herschel~data was not available in the corresponding band. The green solid line is the \cite{Brown2014} spectra templates, plotted as a reference, which are not part of the black-body fits.}
\label{BlackBodyFitsPlotsAll}
\end{center}
\end{figure*}

\begin{table*} [!htb]
\caption{Modified Black-Body Derived Quantities}\label{DustTemperatures}
\centering
\input{Table5_BayesianParameters.tex} 
\end{table*}

\begin{figure*}[hbt] 
\begin{minipage}[l]{0.27\linewidth}
\includegraphics [width=\linewidth]{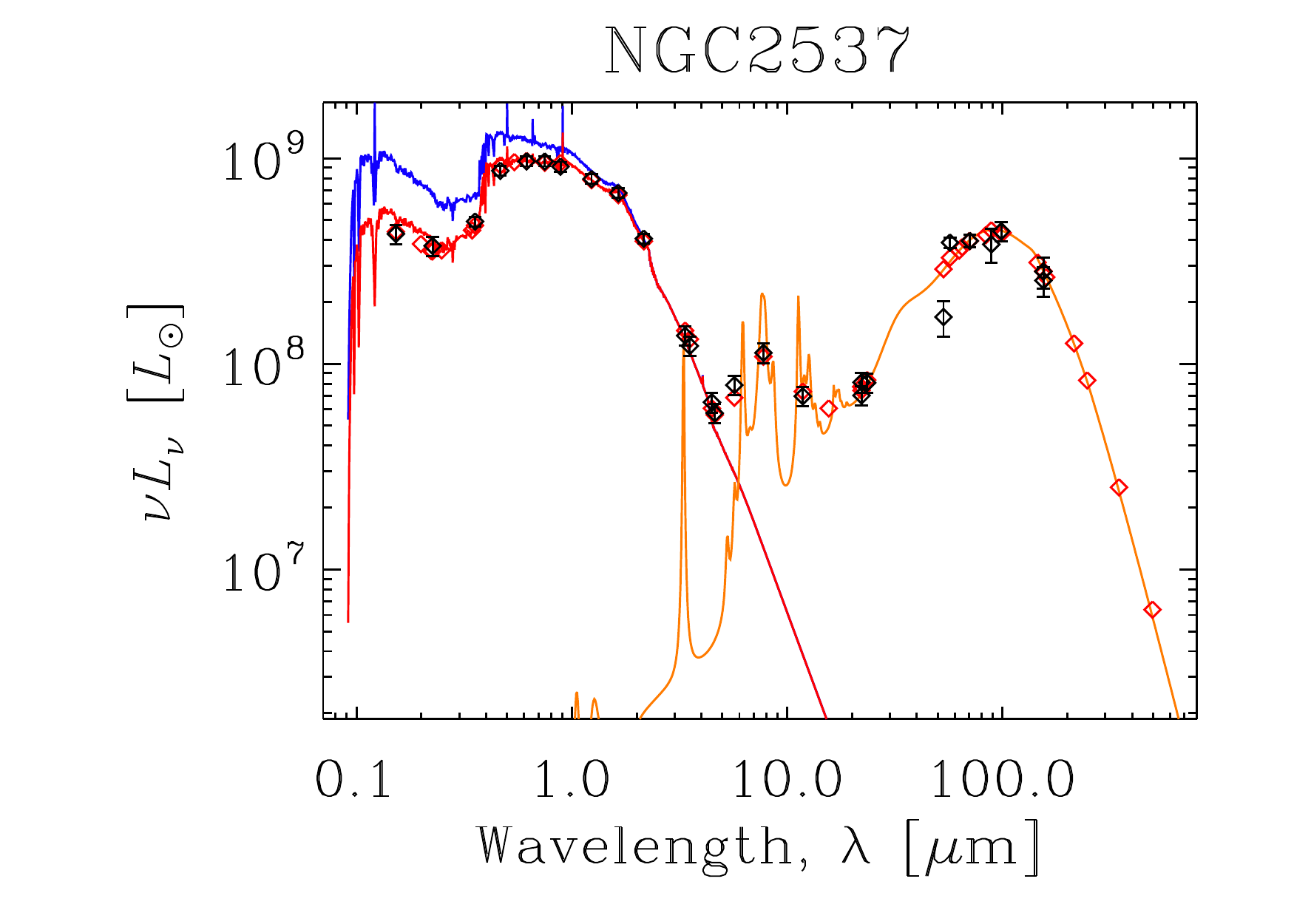} 
\end{minipage} 
\vskip -3.52cm
\hskip 4.2cm
\begin{minipage}[l]{0.27\linewidth}
\includegraphics[width=\linewidth]{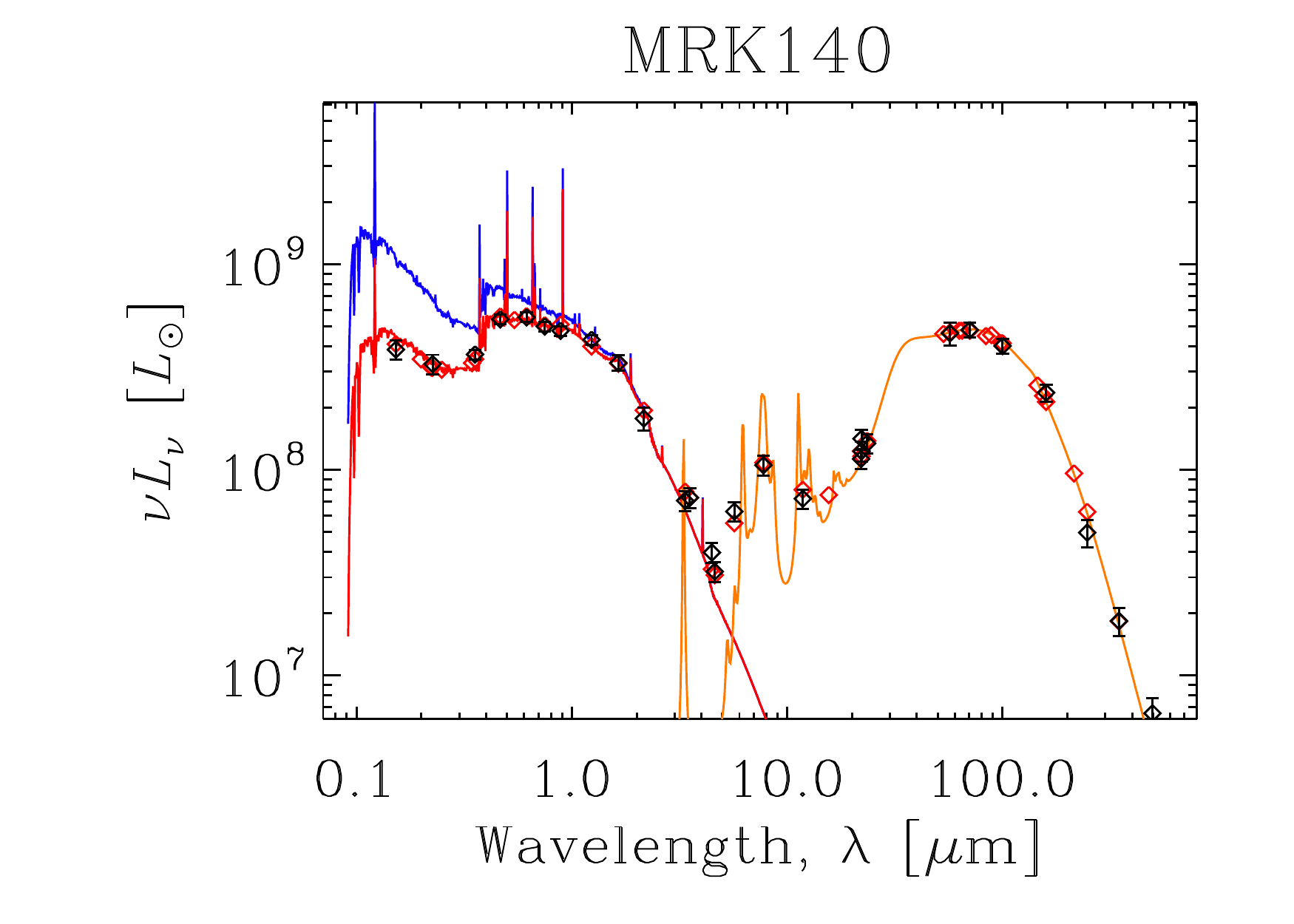}
\end{minipage}
\vskip -3.52cm
\hskip 8.4cm
\begin{minipage}[l]{0.27\linewidth}
\includegraphics[width=\linewidth]{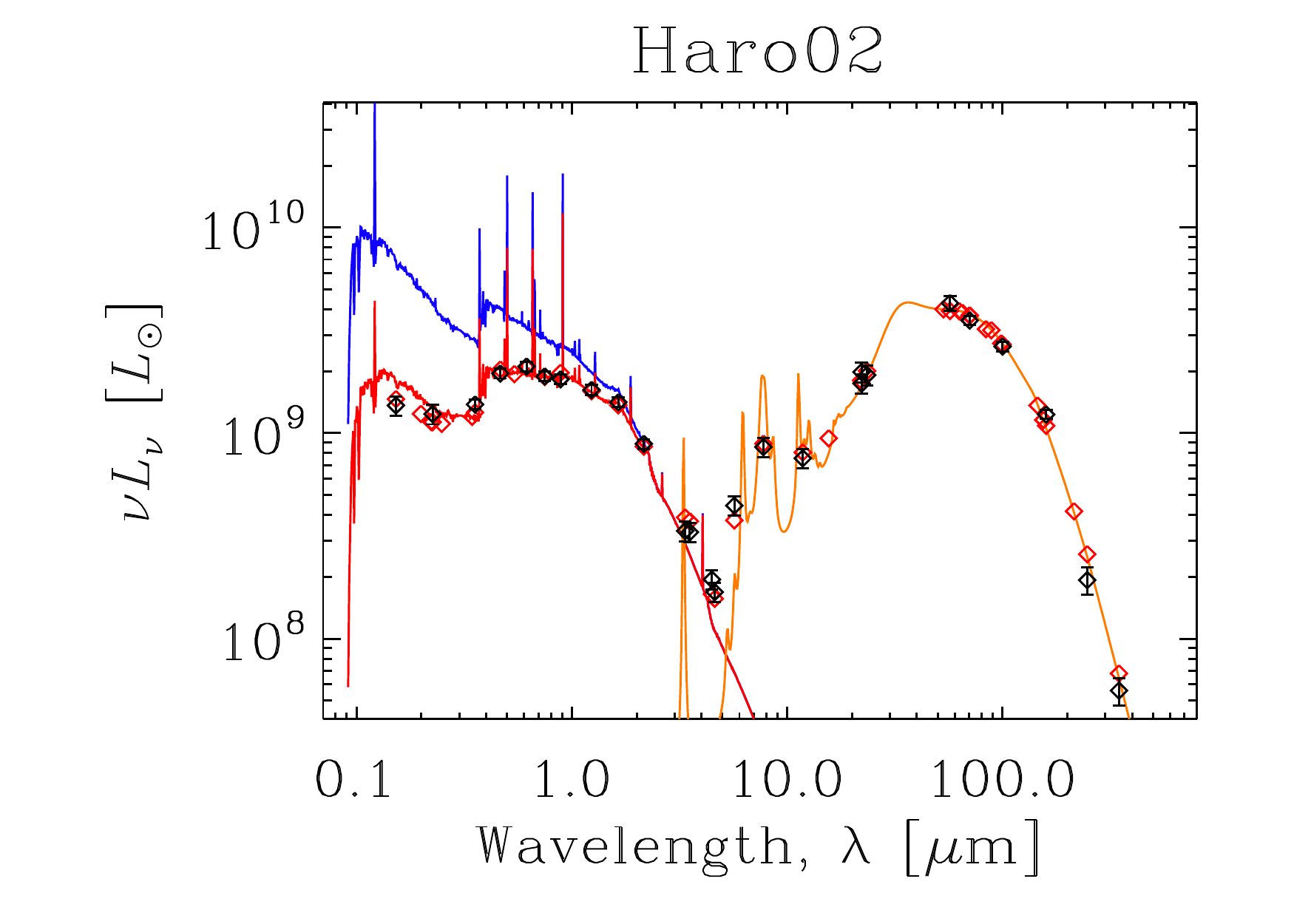}
\end{minipage}
\vskip -3.52cm
\hskip 12.6cm
\begin{minipage}[l]{0.27\linewidth}
\includegraphics[width=\linewidth]{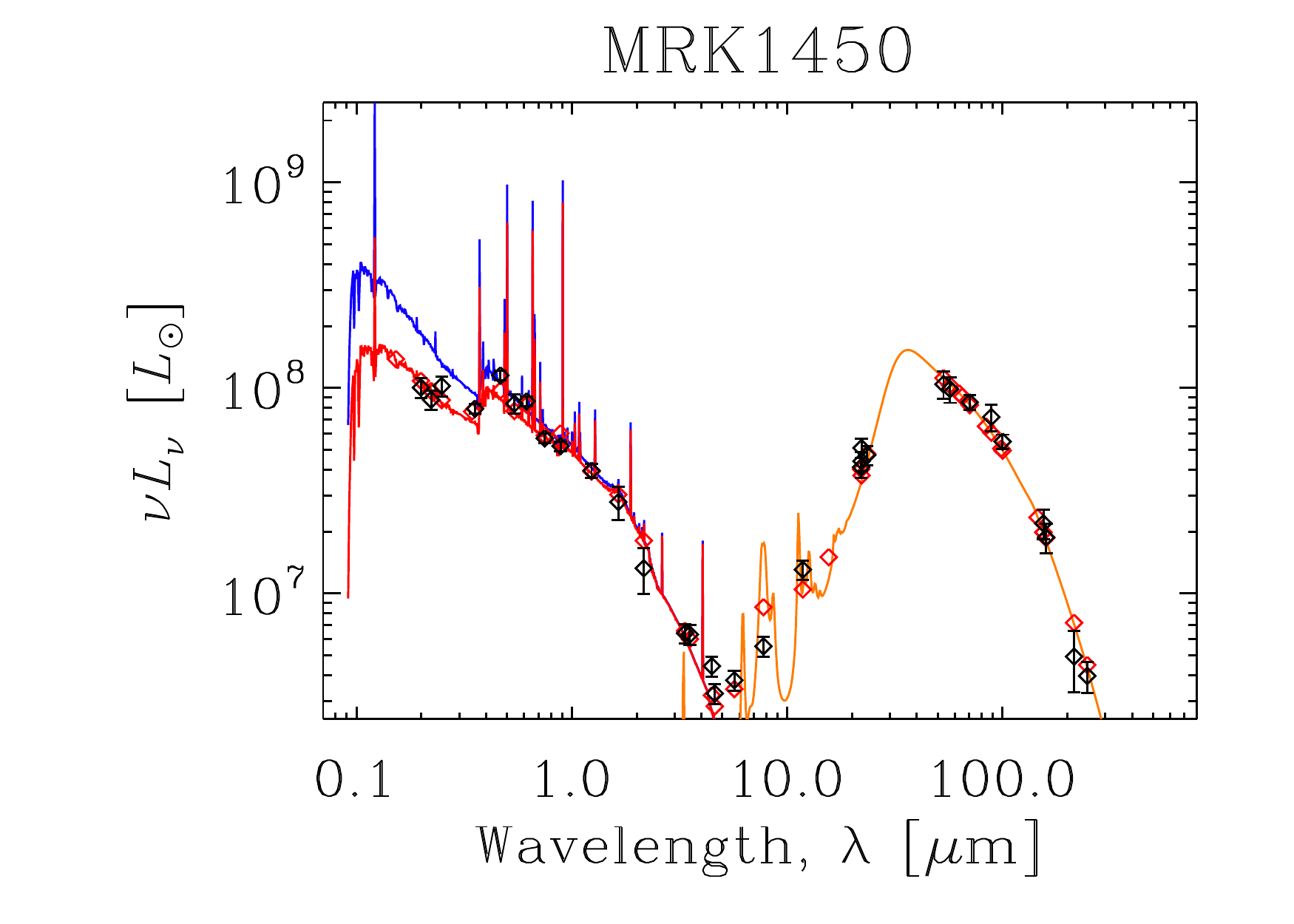}
\end{minipage}
\vskip 0.10cm
\begin{minipage}[l]{0.27\linewidth}
\includegraphics [width=\linewidth]{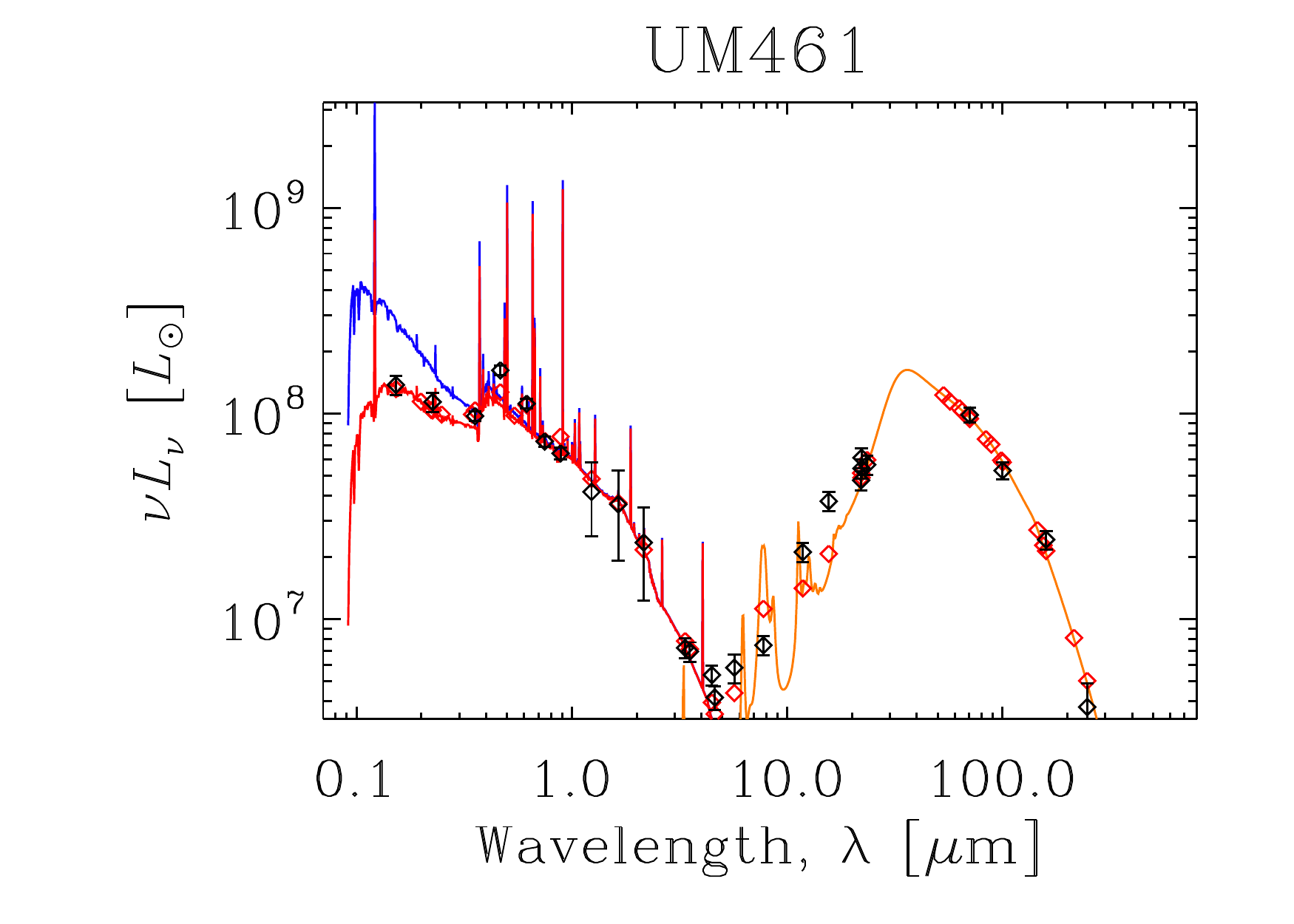} 
\end{minipage} 
\vskip -3.52cm
\hskip 4.2cm
\begin{minipage}[l]{0.27\linewidth}
\includegraphics[width=\linewidth]{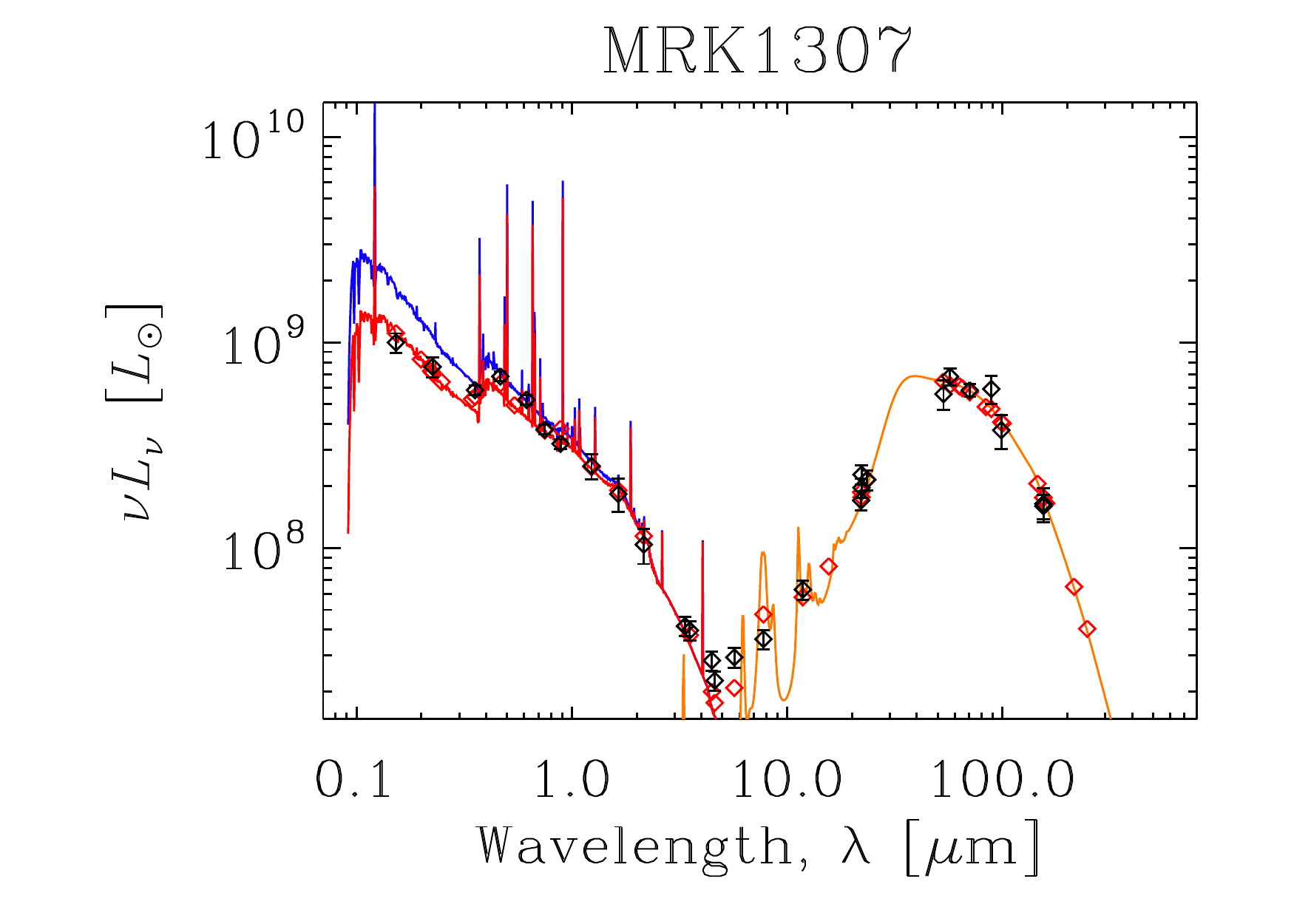}
\end{minipage}
\vskip -3.52cm
\hskip 8.4cm
\begin{minipage}[l]{0.27\linewidth}
\includegraphics[width=\linewidth]{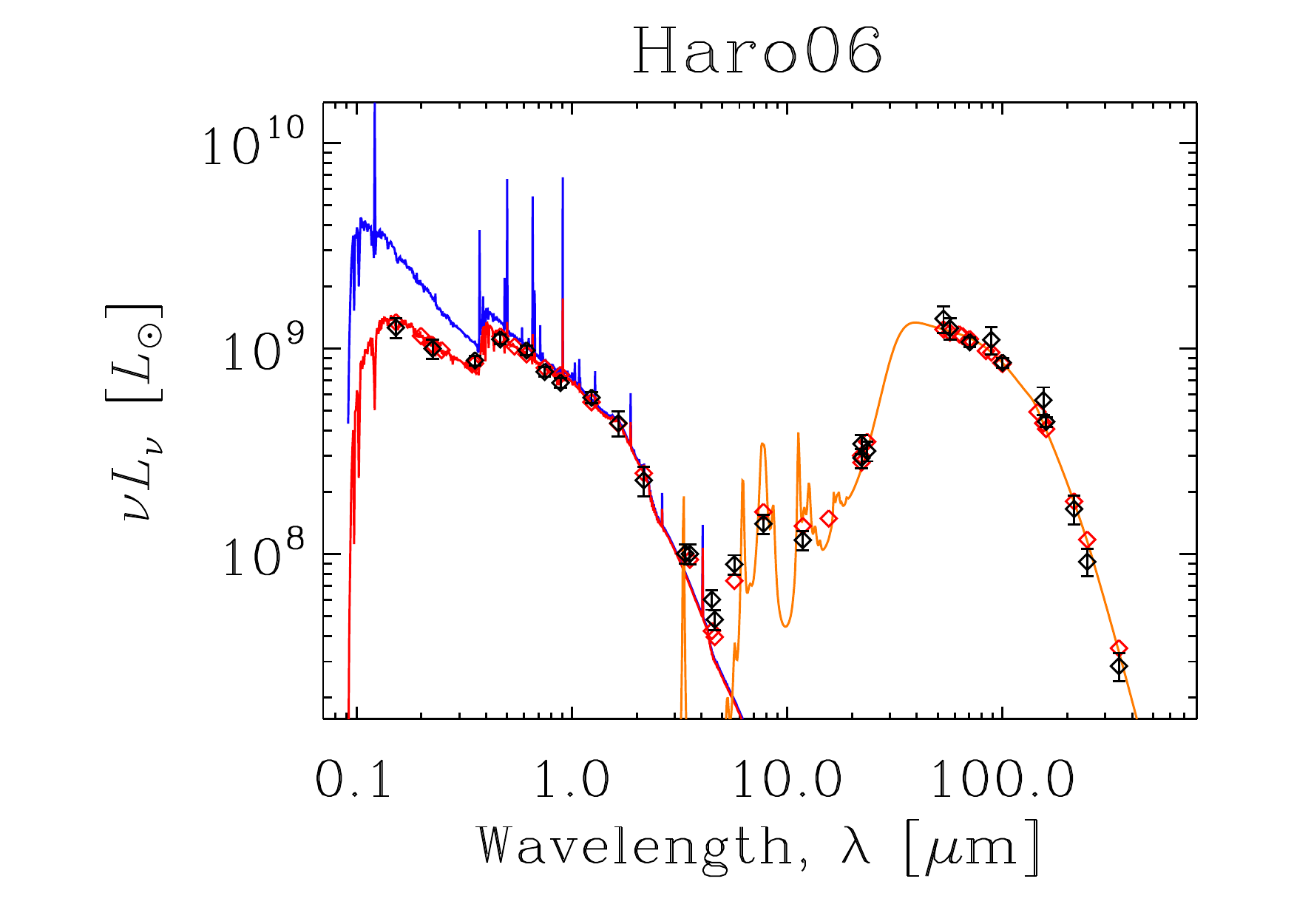}
\end{minipage}
\vskip -3.52cm
\hskip 12.6cm
\begin{minipage}[l]{0.27\linewidth}
\includegraphics[width=\linewidth]{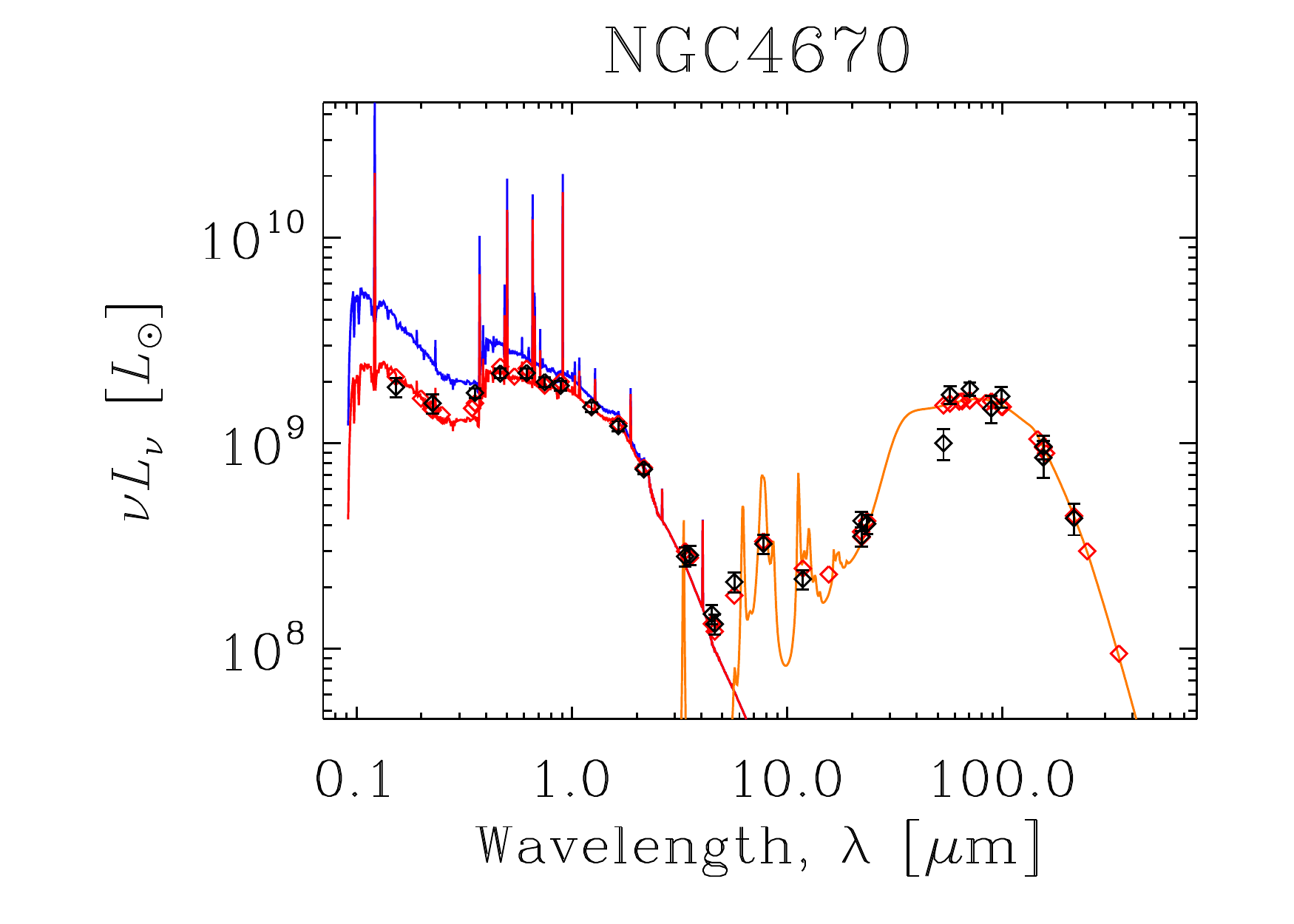}
\end{minipage}
\vskip 0.10cm 
\begin{minipage}[l]{0.27\linewidth}
\includegraphics [width=\linewidth]{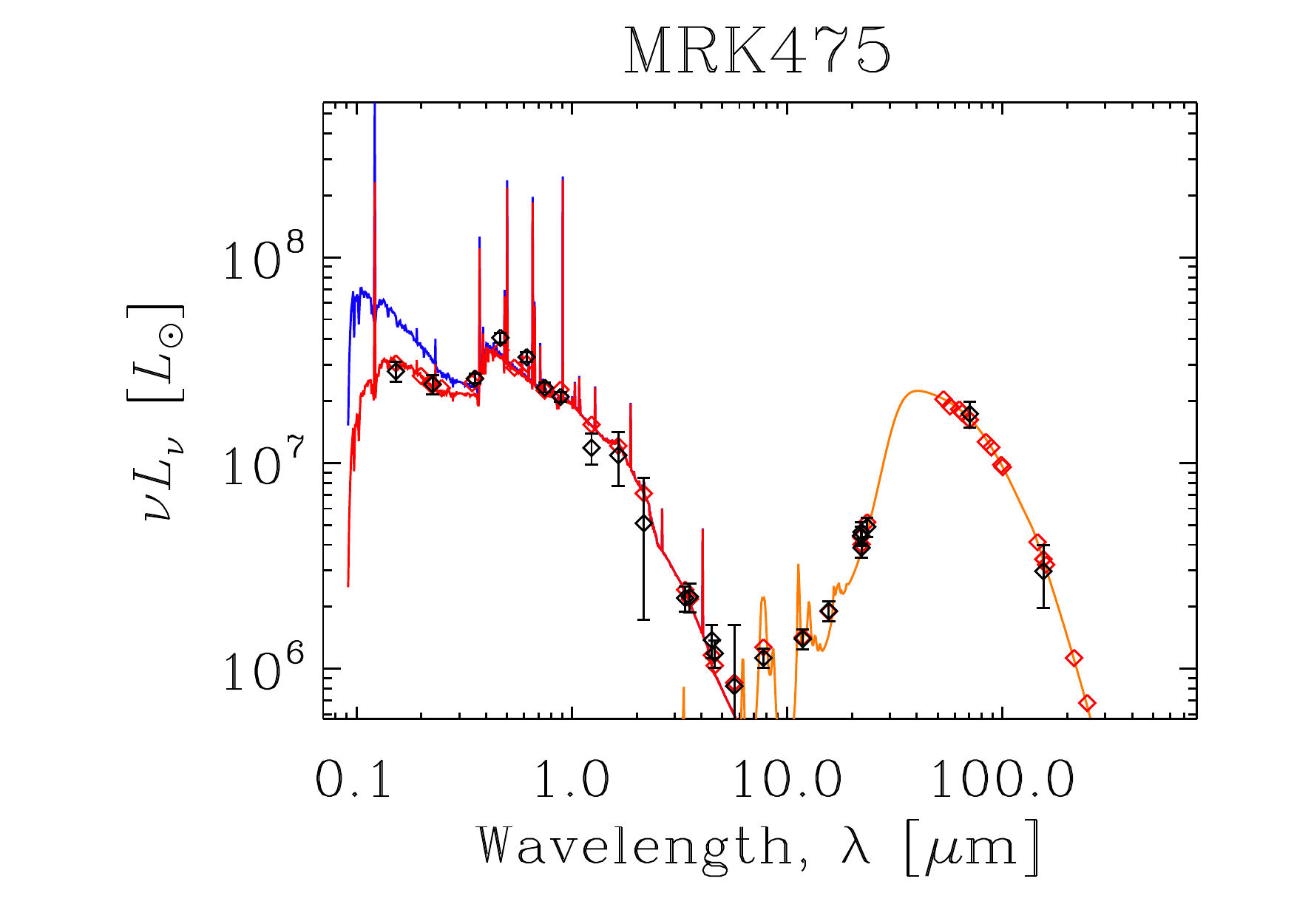} 
\end{minipage} 
\vskip -3.52cm
\hskip 4.2cm
\begin{minipage}[l]{0.27\linewidth}
\includegraphics[width=\linewidth]{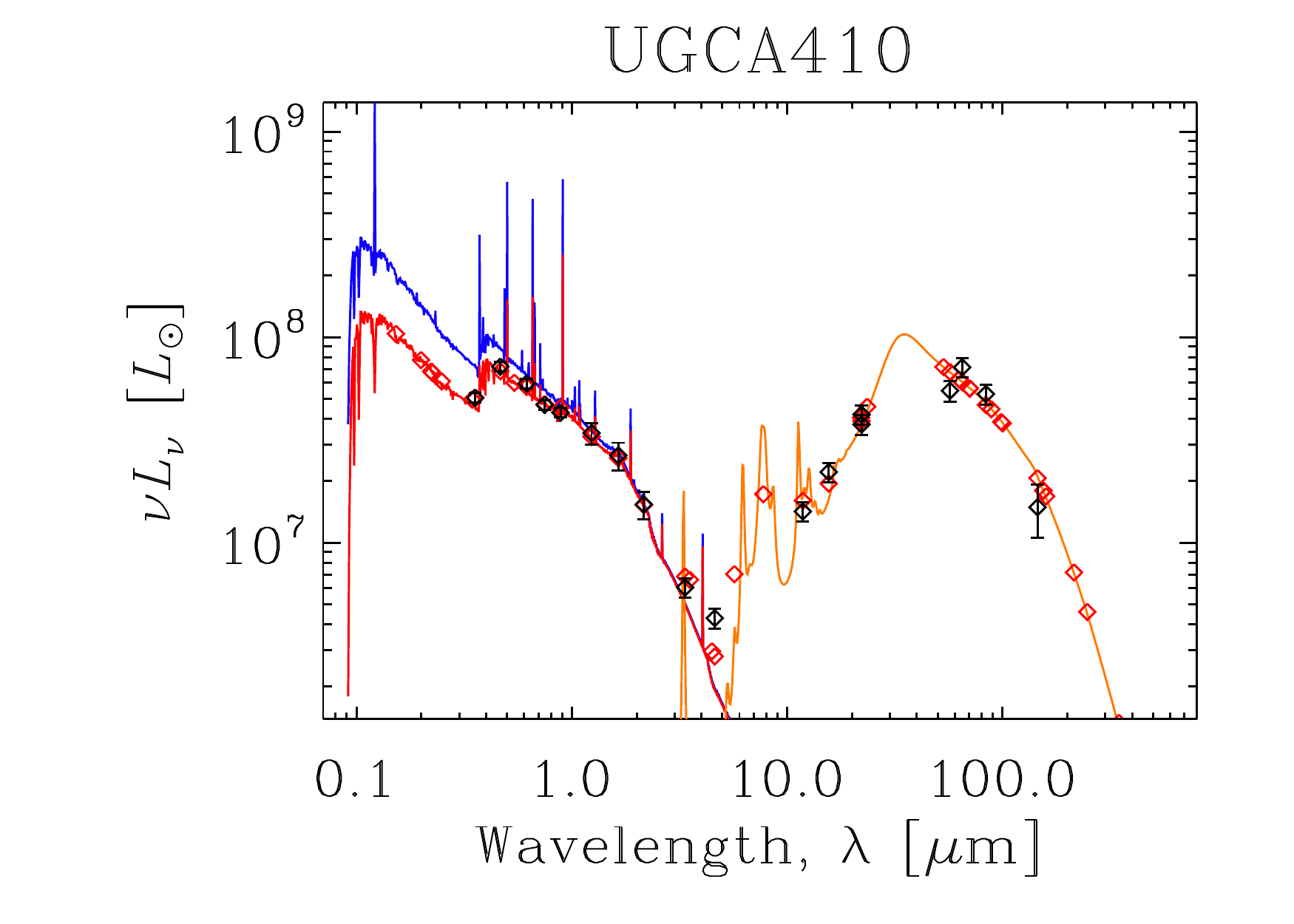}
\end{minipage}
\vskip -3.52cm
\hskip 8.4cm
\begin{minipage}[l]{0.27\linewidth}
\includegraphics[width=\linewidth]{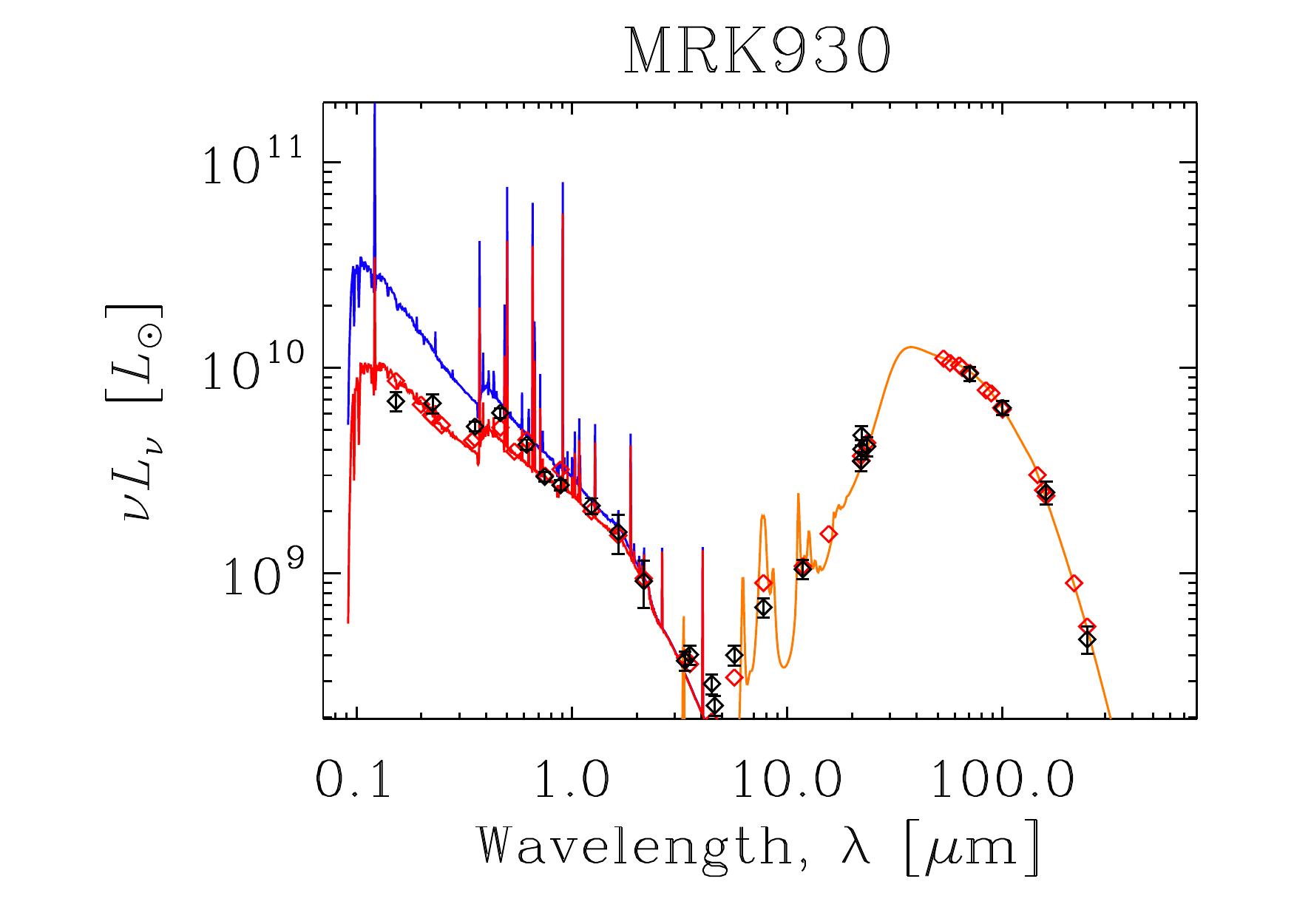}
\end{minipage}
\vskip 0.10cm
\caption{FUV-to-FIR SEDs produced with the \light\ fitting package.  Intrinsic stellar spectra are shown in blue. Attenuated stellar spectra are shown in red. Dust emission spectra are shown in orange. The observed data are shown as black diamonds. Red diamonds are the model SEDs, i.e., the attenuated stellar spectra plus the dust emission spectra convolved with the filter functions. When no observations are available in a band, models are still computed.}
\label{SEDs}   
\end{figure*} 
\begin{table*} [!hbt] 
\centering
\caption{\light-Derived Quantities} \label{LIGHTNINGTable}
\input{Table6_SFR_SFH.tex}
\tablenotetext{1}{The star formation rate averaged over the last 100 Myrs.}
\end{table*}

In Figure \ref{BlackBodyFitsPlotsAll} we present the SEDs, produced by the two-component black-body models for all of our LAHz. In a few cases the warm and the cold components show a double-peaked SED often found for star-forming spiral galaxies. However, in several cases, the warm and cold components blend together and form a broad  single-peaked SED. In the figures we also include the corresponding NIR and MIR spectra from \cite{Brown2014}. These spectra were not part of the fit, but they merge smoothly with the dust SED. In three cases (NGC\,2537, Haro\,02 and NGC\,4670), the MIR spectra suggest that there might be additional flux at wavelengths between $24\,\mu$m and $70\,\mu$m, than what is indicated in the two-component black-body fits.

\subsubsection{Dust temperatures}

We obtained the dust temperatures characterizing the warm and cold components from the modified black-body fits. For comparison we use the Dwarf Galaxy survey presented by \cite{Ruyer2013} which includes local dwarf galaxies, and, in order to provide a comparison with more evolved galaxies, we use a subset of 58 galaxies from the Key Insights on Nearby Galaxies: a Far- Infrared Survey with \herschel\ ({\tt KINGFISH} \citealt{Kennicutt2011}), where we exclude the irregular dwarf galaxies and early type galaxies from the original set of 109 galaxies.

\smallskip

In order to compare our results with the galaxies in the DGS and {\tt KINGFISH} samples, we re-analyzed the FIR photometric data from these surveys using the same modified black body method as for our LAHz galaxies.
For each galaxy we derive the temperatures characterizing the warm and cold components, as well as the temperature corresponding to the peak of the combined SED, defined as:
\begin{equation}
T_{\rm peak} = \frac{b}{\lambda_{\rm peak}}\,,
\end{equation}

where $b$ is the Wien's displacement constant.
All the galaxies were characterized using two dust components. The average dust temperatures for the sample of LAHz are $T_{\rm d,\,warm}\sim124\,$K, and $T_{\rm d,\,cold}\sim42\,$K. This is significantly higher than for star-forming disk galaxies in the {\tt KINGFISH} sample, with $T_{\rm d,\,warm}\sim68$\,K and $T_{\rm d,\,cold}\sim23$, and higher than for the DGS sample with $T_{\rm d,\,warm}\sim74$\,K and $T_{\rm d,\,cold}\sim24$. The range of temperatures for the DGS survey for the cold component is, however, quite large, ranging from $ 18-41$\,K (\citealt{Ruyer2013}). The results are shown in Figure~\ref{TempsComparison}.

\subsection{\light\ models}

As described in Sect.\ref{sec:light}, the \light\ models fit the whole UV-FIR SED, while self-consistently applying an extinction correction. The FIR is modeled using \cite{DraineLi2007} templates. The resulting SEDs for our LAHz galaxies
are shown in  Figure~\ref{SEDs}.
The black data points represent the observed SED; the blue line is the spectra without dust attenuation; the red line is the SED with attenuation; the orange line is the best-fit dust emission. In Table \ref{LIGHTNINGTable} we show luminosities, dust masses, SFR averaged over the last 100 Myrs and specific star formation rates (sSFRs) derived from the \light\ models.

\subsection{Far-infrared luminosity and dust mass} \label{LIRandMasses}

We derive dust masses and $L_{\rm IR}$ using the SEDs produced with the two different modeling methods described in Sect.~\ref{Sec:Modeling}. The dust masses and $L_{\rm IR}$ from the mBB models and the \light\ method are presented in Table~\ref{DustTemperatures} and Table~\ref{LIGHTNINGTable} respectively.
The $L_{\rm IR}$ is derived by integrating the SED from $8\,\mu$m to $1000\,\mu$m. The LAHz galaxies luminosities cover a wide range of $\log(L_{\rm IR}/L_{\odot})=\,[7.4,\,10.2]$ for the black-body models, similarly, the \light\ method
luminosities range between $\log(L_{\rm IR}/L_{\odot})$=$\,[7.5,\,10.3]$. 

\smallskip

The derived dust masses using the black-body models are in the range of: $\log{(M_{\rm dust}/M_{\odot})}=\,[3.1,\,6.3]$. As expected the cold dust component dominates the total dust mass. We also derive the dust mass from the Draine \& Li best fits in the \light\ models: $\log{(M_{\rm dust}/M_{\odot})}=\,[3.2,\,6.3]$. Some of the galaxies in our sample have published dust mass values, derived using a subset of the photometric data used here (e.g. \citealt{Engelbracht2008}; \citealt{Ruyer2013}), which agree very well with the results presented here. 
For the discussion presented in the rest of this paper, we use the dust masses derived with the mBB method, since this allows the comparison with dust masses derived for surveys samples like the {\tt KINGFISH} and DGS using the same methodology. 

\smallskip

We compare the dust masses and FIR luminosities for our LAHz galaxies, derived with the modified black body and \light\ methods. The latter method uses the \cite{DraineLi2007} templates
to model the FIR SED.
The dust masses derived from the  {\it mBB} and the Draine \& Li models, show  dispersion of $\sim$0.3dex, but shows no systematic difference. The FIR luminosities,
however, do show a systematic offset, where the \cite{DraineLi2007} based models give $\sim$20\% higher L$_{\rm FIR}$ than those derived by integrating over the modified black body models. This is expected since the \cite{DraineLi2007}
models include PAH emission in the near-infrared, and better account for the near- to mid-IR emission than the black body models.
In the rest of the paper we will use the {\it mBB}-derived dust masses for comparison with the DGS and {\tt KINGFISH} surveys. And the \light\ FIR luminosities for discussions and conclusions.

\subsection{Star formation histories}

\begin{figure*}[!htb] 
\begin{minipage}[l]{\linewidth}
\includegraphics [width=\linewidth]{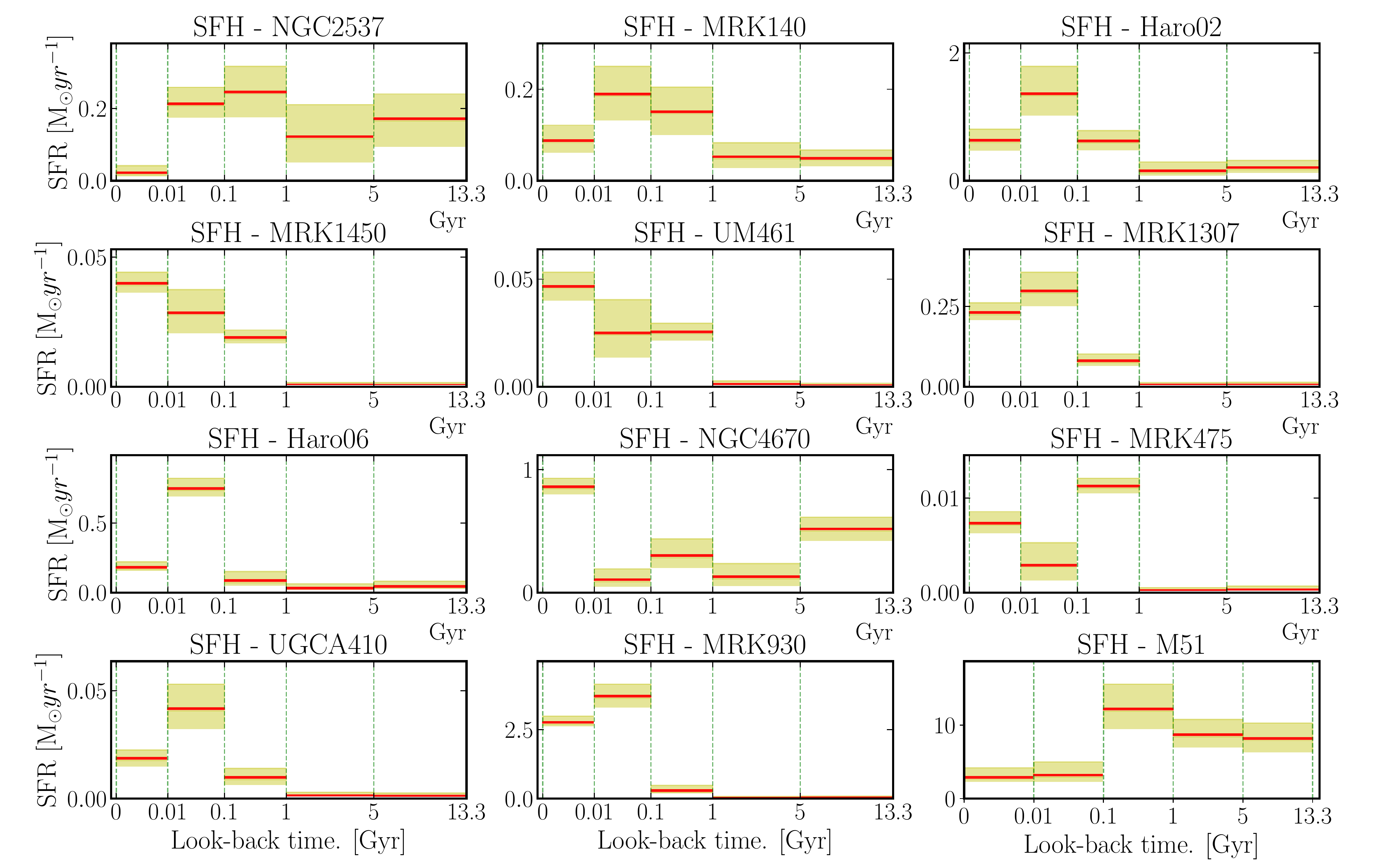}
\caption{Star formation histories for our sample. The SFHs are divided into five time bins: 0 to 10\,Myr, 10 to 100\,Myr, 100\,Myr to 1\,Gyr, 1 to 5\,Gyr and 5 to 13.3\,Gyr. 
The solid lines denote the median values from the MCMC chain and dotted lines show the 16th and 84th percentiles, enclosing the hashed areas. For comparison the last frame includes the SFH of M51, derived by  \cite{Eufrasio2017}, following similar approach.}\label{SFH} 
\end{minipage}
\end{figure*}

 In Fig.~\ref{SFH} we show the SFH for the LAHz galaxies in our sample. For comparison we also show the SFH of M51 from \cite{Eufrasio2017}. The SFH was derived using the \light\ fitting package, and it is presented in 5 look-back time bins:
 $0-10$~Myrs, $10-100$~Myr, $0.1-1$~Gyr, $1-5$~Gyr and $5-13.3$~Gyr. The \light\ modeling treats the SFR as constant over each time bin.
The solid red line denote the median SFR for each time bin from the simulations and the yellow region denotes the 16th and 84th percentiles, enclosing the 68\% confidence intervals.
The SFHs of the LAHz are for the most part very different from those of more massive galaxies. For instance, the SFH of M51, a large spiral galaxy gravitationally interacting with a neighbor galaxy, shows a more complex SFH, with a significant star formation contribution in bins corresponding to ages older than 5~Gyr.

\smallskip

For the LAHz we recover SFHs consistent with null SFR contribution at ages older than 1~Gyr for 6 galaxies: Mrk~1450, UM~461, Mrk~1307, Mrk~475, UGCA 410 and Mrk~930. This suggests that they are the youngest galaxies in our sample. 
Two of the galaxies, NGC\,2537 and NGC\,4670, show significant star formation activity in all five time bins, indicating the presence of an old stellar population. Hence, despite being designated as Blue Compact Dwarf Galaxies, the stellar mass of 
these two galaxies is dominated by old stars. The situation is less clear for three of the galaxies in our sample, Mrk\,140, Haro\,02 and Haro\,06, where the \light\ results indicate that it is probable that they had a low level of star formation activity
at times $\gtrsim$1\,Gyr (see Figure~\ref{SFH}).

\smallskip

We consider the six LAHz galaxies that have SFHs consistent with being younger than $\sim$1\,Gyr as our prime candidates for being local analogs. NGC\,2537 and NGC\,4670 are clearly not young galaxies. Mrk\,140, Haro\,02 and Haro\,06
are likely to have an old stellar population, although Haro\,06 is a borderline candidate for being a local analog.

\section{Discussion}\label{Sec:Discussion}

\subsection{Characteristic dust temperatures}

When fitting a two-component modified black body curve the our LAHz galaxies, the Dwarf Galaxy Survey and spiral galaxies in the {\tt KINGFISH} sample, we find that the DGS and LAHz galaxies show a higher $T_{\rm peak}$ than the
more massive {\tt KINGFISH} galaxies (Figure~\ref{TempsComparison}). While the average $T_{\rm peak}$ is 33\,K and 35\,K for the LAHz and DGS samples, respectively, the {\tt KINGFISH} galaxies have an average of $T_{\rm peak} = 23$\,K.
When we compare the warm/cold dust component separately, we see that the LAHz are characterized by higher dust temperatures than both the {\tt KINGFISH} and the DGS galaxies. These results are consistent with a general trend of warmer
dust in metal-poor galaxies compared with the more metal-rich and more massive counterparts (e.g. \citealt{Ruyer2013}) as well as for high-$z$ galaxies (e.g. \citealt{Faisst2017} and \citealt{Sommovigo2020}). 

\begin{figure*}[htb] 
\begin{center}
\includegraphics [width=\linewidth]{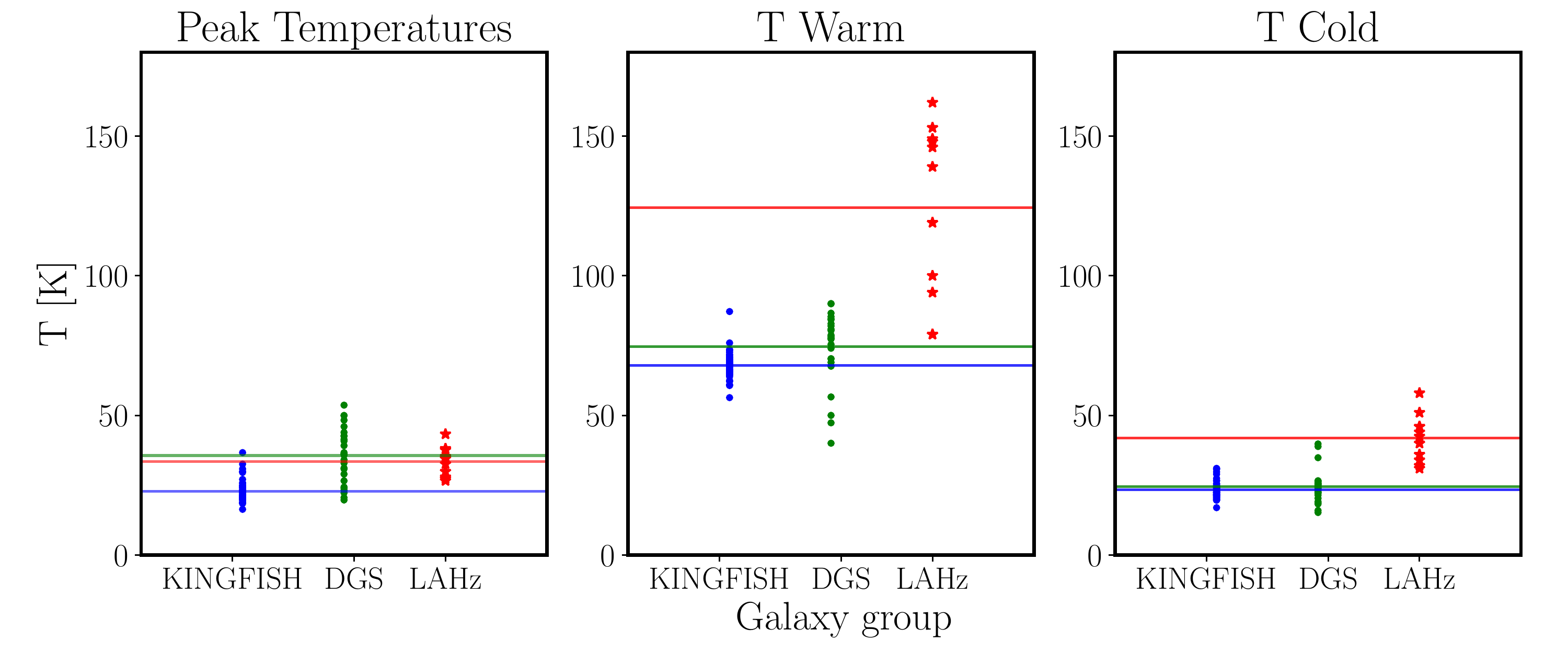}
\end{center}
\caption{Comparison of peak, warm, and cold temperatures characterizing the dust. Solid lines indicate average temperatures; the dots are the individual values for each galaxy. {\tt KINGFISH} galaxies are shown in blue, DGS in green, and the LAHz candidates in red. All temperatures were determined using modified black-body procedure (see section \ref{bbmodel}).}
\label{TempsComparison}   
\end{figure*} 

\subsection{Dust-to-stellar mass ratio}
In Figure~\ref{MdustMstarLIRZ}, (top-left frame), we show the dust-to-stellar mass ratio as a function of stellar mass for our sample, together with those of the DGS, and {\tt KINGFISH} spiral galaxies. While the {\tt KINGFISH} galaxies seem
to have a weak dependence on stellar mass, no such correlation is apparent for our LAHz galaxies or for galaxies in the DGS sample. In fact, the dust-to-stellar mass ratio for the LAHz seems to be approximately constant, with a median ratio $\log(M_{\rm dust}$/$M_{\star})=(8.5 \pm 5) \times 10^{-4}$. A similar argument can be made for the DGS sample, but with a slightly lower average dust to stellar mass ratio.
The dashed line shown in Figure \ref{MdustMstarLIRZ} shows the anti-correlation between dust-to-stellar mass ratio and stellar mass found by \cite{Clemens2013}. This anti-correlation was obtained for galaxies from the Planck Survey and the fit is dominated by galaxies in the stellar mass range $\log{M_{*}/M_{\odot}} \sim 10-11$. The anti-correlation is surprising in view of the fact that the gas-phase metallicity scales with stellar mass (e.g. \citealt{Tremonti2004}) and one would expect a lower
$M_{\rm dust}/M_{\star}$ ratio for low-mass galaxies. The \cite{Clemens2013} relation has been interpreted as being due to a rapidly increasing gas fraction for lower-mass galaxies, more than offsetting the metallicity effect. However, the results shown here, both for our sample and that of DGS, show that this relation cannot be extended to low-mass galaxies.

\smallskip
 
The mass-metallicity relation is established down to $\log{M_*/M_{\odot}} \sim 8.5$. Below this stellar mass, depletion of metals through galactic winds starts to become important (e.g. \citealt{Tremonti2004}). Our LAHz galaxies have $\log{M_{*}/M_{\odot}}$ in the range $7.0 - 9.4$, straddling the $\log{M_{*}/M_{\odot}} = 8.5$ value. In Figure~\ref{MdustMstarLIRZ}, top-right panel, we show the relation between the dust-to-stellar mass ratio as function of metallicity. In the figure we also include the DGS and {\tt KINGFISH} spiral samples. Based on the latter two samples, \cite{Ruyer2013} defined a correlation between the dust-to-stellar mass ratio and gas-phase metallicity, shown as a dashed line in Figure~\ref{MdustMstarLIRZ}. We find no obvious correlation between the dust-to-stellar mass ratio and metallicity for our LAHz sample, and the overall dispersion of dust-to-stellar mass ratios is large. As discussed in \cite{Ruyer2013}, this could at least in part be due to observational uncertainties. However, for the metal-poor, and low-mass galaxies, this could also be due to the impact of individual star formation histories, where the interstellar gas can be promptly enriched through recent star formation, or depleted due to galactic winds. Evidence for both effects have been seen in this type of galaxies; \cite{Lagos2018} found spatial variations in the gas-phase metallicity in UM\,461, attributed to recent infall of low-metallicity gas, and Haro\,02 shows clear evidence of a galactic gas outflow (e.g. \citealt{Meier2001}; \citealt{Beck2020}).

\smallskip
 
\begin{figure*}[!htb] 
\begin{minipage}[l]{0.99\linewidth}
\includegraphics [width=\linewidth]{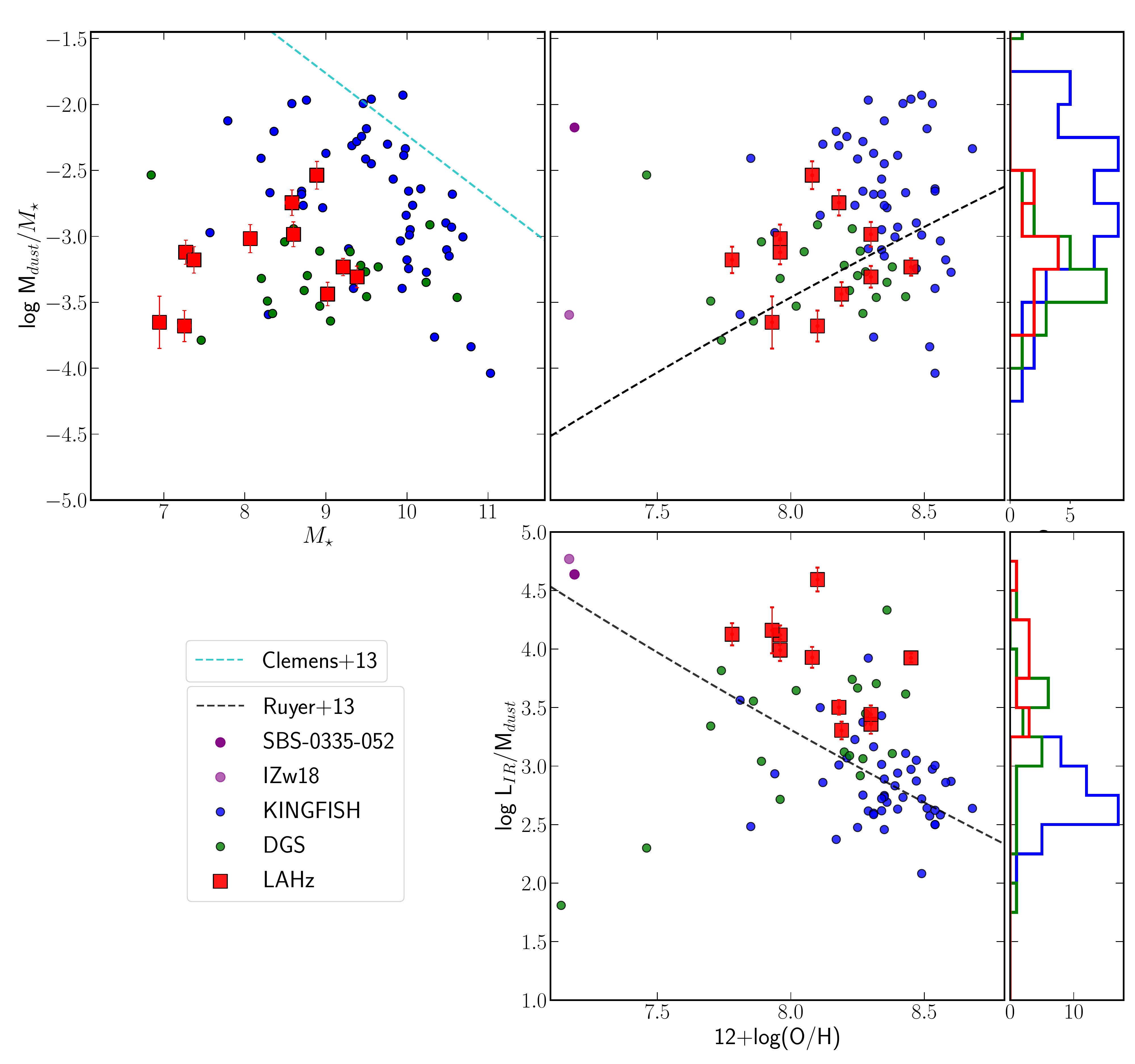}
\end{minipage}
\caption{{\bf Top left:} The dust to stellar mass ratio as a function of stellar mass.  The magenta dashed line denotes the relation from \cite{Clemens2013}, based on more massive galaxies.
{\bf Top right:} Dust to stellar mass ratio as a function of metallicity. The power-law defined by \cite{Ruyer2013} is shown as a dashed line. For comparison we include the extremely low-metallicity galaxies SBS-$0335-052$ and IZw18 (purple circles; e.g. \cite{Hunt2014}).
{\bf Bottom right:} $L_{\rm IR}$/$M_{\rm dust}$ as a function of metallicity. The dashed line corresponds to the relation found by \cite{Ruyer2013}. We include SBS-$0335-052$ and IZw18 as purple circles.,
The histograms on the right show the distribution of each group.
In all three panels: red squares are the LAHz candidates; green circles are the DGS sample; blue circles represent the {\tt KINGFISH} spiral sample.
}\label{MdustMstarLIRZ}   
\end{figure*}
 
For comparison we also include two of the lowest metallicity galaxies known, SBS\,0335-052E and IZw18 in Figure~\ref{MdustMstarLIRZ}.  Both of these galaxies have very low gas-phase metallicity, $12+\log{O/H} \sim 7.2$, but even at these extremely low metallicities they still show dust continuum emission (SBS\,0335-052E: \cite{Hunt2014}; IZw18: \cite{Wu2007}. These two particularly low-metallicity galaxies were studied by \cite{Hunt2014}, and found to have dust-to-stellar mass ratios within the range of values found for the galaxies presented in Figure~\ref{MdustMstarLIRZ}. The dust-to-stellar mass ratio of IZw18 is $\log({M_{\rm dust}/M_{\star}})=-3.60$, which is consistent with the LAHz average of $\log({M_{\rm dust}/M_{\star}})=-3.17$. SBS\,0335-052E, however, has $\log({M_{\rm dust}/M_{\odot}})\,\approx -2.15$, which is significantly higher than the LAHz and DGS galaxies, comparable with metal-rich spiral galaxies. While this wide disparity among the most metal-poor galaxies could be due to observational uncertainties, it could also be an effect of the above mentioned prompt enrichment and galactic outflow/inflow processes affecting low-mass galaxies.

\smallskip

The bottom-right panel of Figure~\ref{MdustMstarLIRZ} shows the $L_{\rm IR}/M_{\rm dust}$ ratio as a function of metallicity for the LAHz galaxies (red squares). For comparison we also show the DGS and {\tt KINGFISH} galaxies. The distribution of $L_{\rm IR}/M_{\rm dust}$ is indicated  on the side for each group. The $L_{\rm IR}/M_{\rm dust}$ of the LAHz galaxies is distributed in the same range as the DGS (\citealt{Ruyer2013}). This ratio tends to increase for lower metallicities, and when we examine the left region in the range of metallicities, the LAHz exhibit higher $L_{\rm IR}/M_{\rm dust}$ ratio that the DGS or {\tt KINGFISH} galaxies for the corresponding metallicity. This means that the dust grains in the LAHz sample are more efficient, on average, at radiating far-infrared photons. It is not possible with the present data to determine whether this is due to different grain properties, or is simply an effect of the higher dust temperatures characterizing the dust emission in these galaxies. As shown in Figure~\ref{MdustMstarLIRZ}, this trend extends to the extremely metal-poor galaxies SBS\,0335-052E and IZw18, covering more than an order of magnitude in both metallicity and $L_{\rm IR}/M_{\rm dust}$. These two galaxies are dominated by warm dust; their single-component dust SEDs are characterized by $T_{\rm dust} = 59$ and $67$\,K, respectively (\citealt{Hunt2014}). While these temperatures are high, they are within the range of the temperatures characterizing our LAHz galaxies.

\begin{figure}[!htb]
\includegraphics [width=\linewidth]{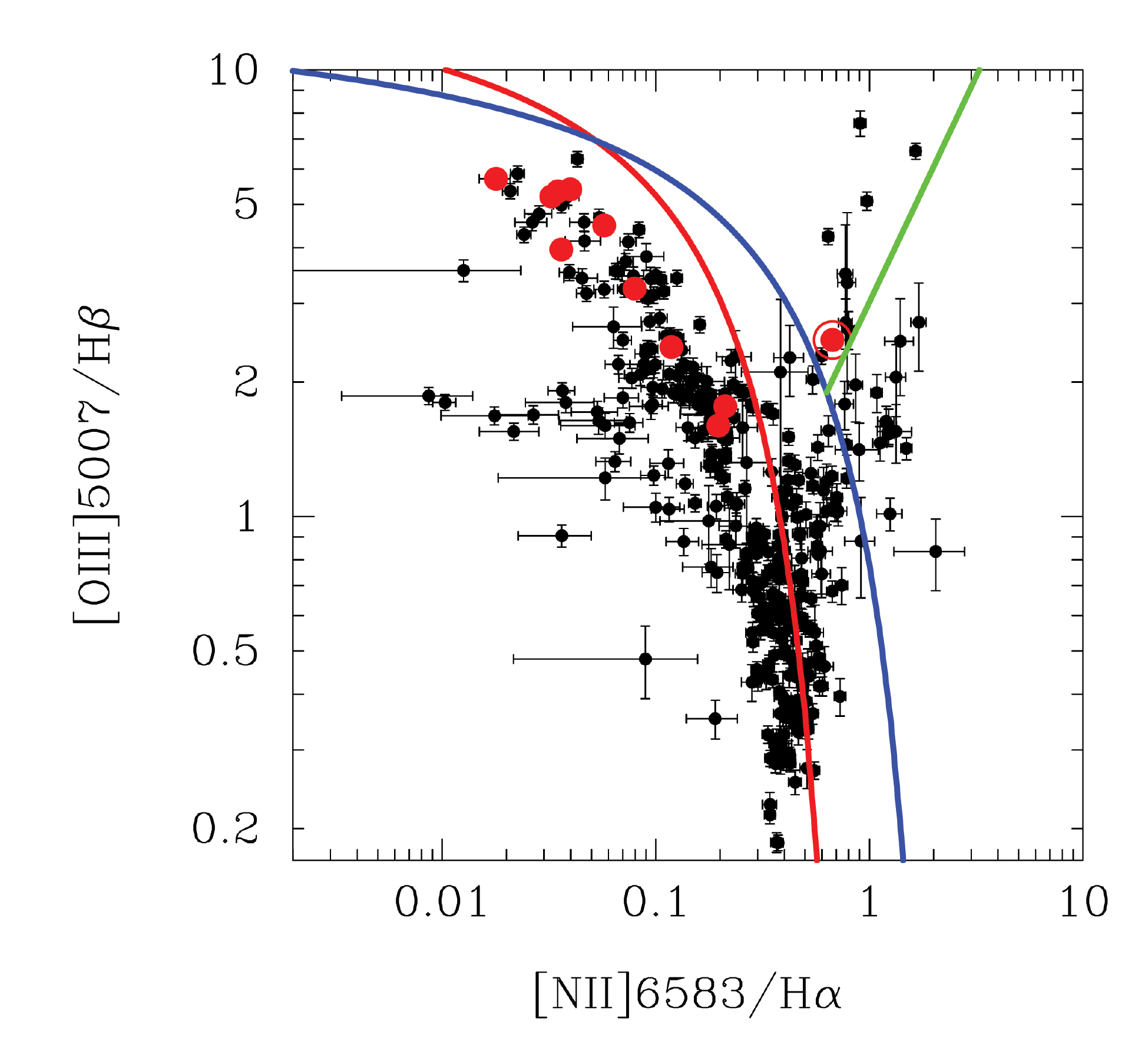}
\caption{BPT diagram (\citealt{Baldwin1981}) for nearby galaxies in the \cite{Moustakas2006} sample. This sample contains all eleven of our Local Analog candidates, marked as red circles. The blue line depict the "maximum starburst line" (\citealt{Kewley2001}) and is an upper limit for excitation caused by star formation. The red line is an alternate definition based on Sloan Digital Sky Survey galaxies (\citealt{Kauffmann2003}). The green line divides Seyferts and LINERS (\citealt{CidFernandes2010}). }\label{fig:bpt}

\end{figure}


\subsection{Star Formation vs. AGN}

The high temperatures characterizing the warm dust component of the LAHz galaxies, as well as the high MIR fluxes relative to the peak far-infrared flux of several of the LAHz galaxies, could potentially be due to the presence of a non-stellar
heating source, such as an active galactic nuclei (AGN).  AGNs in low-mass galaxies are more difficult to detect than their counterparts in massive galaxies. None of our LAHz galaxies are, however, believed to harbor an active AGN. \cite{Reines2020} used the VLA to show that previously identified radio sources that could potentially be AGNs in Haro\,02 and Mrk\,1307 are most likely associated with star formation activity. 

\smallskip

We explore the possibility of AGN, LINER or Seyfert activity in the LAHz galaxies by plotting them in a BPT diagram (\citealt{Baldwin1981}). This diagnostic uses the H$\alpha/N[II]$ ratio together with the [OIII]5007/H$\beta$ ratio to distinguish between star formation and AGN activity. The spectral line observations presented in \cite{Moustakas2006} includes all of our LAHz galaxies, and the BPT diagram of the galaxies in \cite{Moustakas2006} is shown in Figure~\ref{fig:bpt}, with our LAHz galaxies marked as red circles. All LAHz galaxies, except Mrk\,140 falls in the star formation part of the diagram. Mrk\,140 falls close to the line depicting maximum star formation activity as defined by \cite{Kewley2001}.
Since there are no other indication suggesting that Mrk\,140 harbors an AGN, we will treat it as a pure star-forming system.

\subsection{Star Formation Rates }
\begin{figure*}[!hbt] 
\begin{minipage}[l]{0.52\linewidth}
\includegraphics [width=\linewidth]{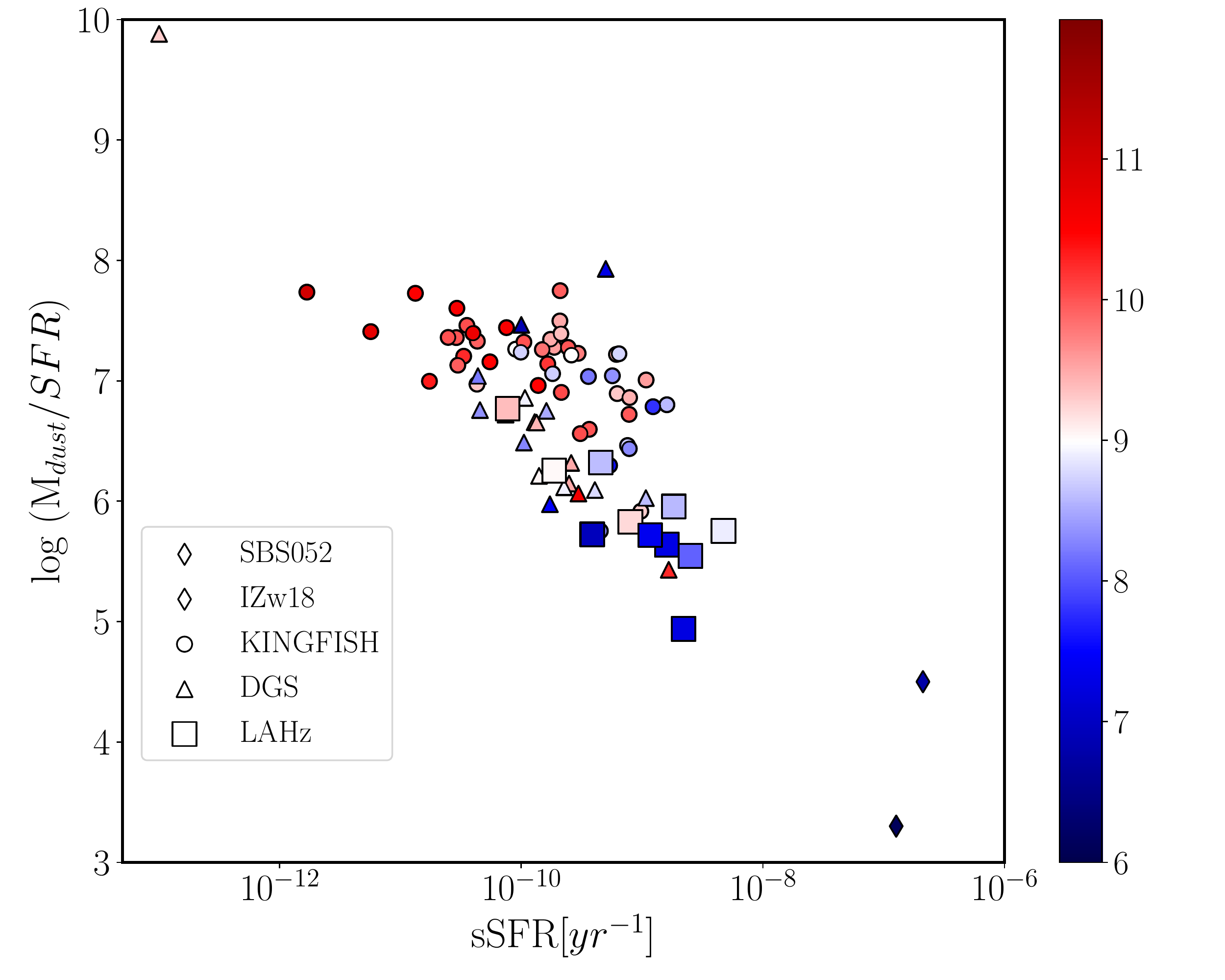}
\end{minipage}
\vskip -7.5cm
\hskip 9.00cm
\begin{minipage}[r]{0.52\linewidth}
\includegraphics [width=\linewidth]{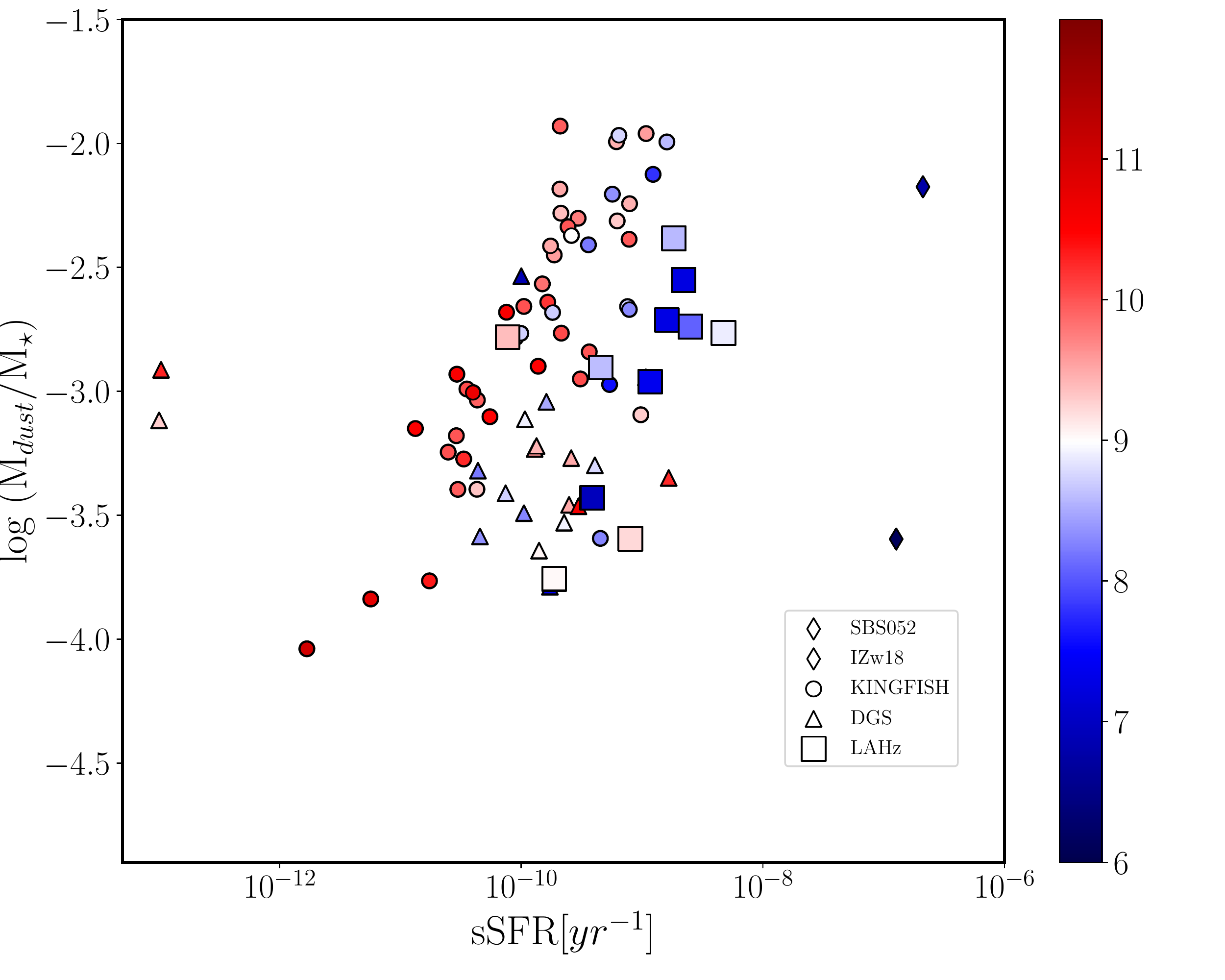}
\end{minipage}
\caption{{\bf Left:} The ratio of dust mass and star formation rate vs specific star formation rate. {\bf Right:} The ratio of dust and stellar mass vs specific star formation rate. In both frames, the LAHz candidates are shown as squares. For reference we include the {\tt KINGFISH} galaxies as circles (\citealt{Kennicutt2011}), and the DGS galaxies as triangles (\citealt{Ruyer2013}). The two extreme low metallicity galaxies IZw18 and SBS-$0335-052$ (\citealt{Hunt2014}) are shown as diamonds. The color indicates stellar mass as shown in the right bar. More massive galaxies are red, less massive are blue.}\label{Mdust_sSFR}   
\end{figure*}

In Figure~\ref{Mdust_sSFR} we show the correlations between $M_{\rm dust}/SFR$ and $M_{\rm dust}/M_{\star}$ as a function of the specific star formation rate (sSFR\,$=\,SFR/M_{\star}$).  As shown in \cite{DaCunha2010}, the dust-to-stellar
mass ratio strongly correlates with the sSFR, confirming that dust mass and star formation are tightly related. In Figure.~\ref{Mdust_sSFR} we color code the stellar mass of each galaxy.
More massive galaxies exhibit a lower sSFR, and higher $M_{dust}/SFR$ ratio, contrary the lower-mass galaxies, which show higher specific SFRs, tracing a clear decreasing relation between $\log{(M_{dust}/SFR)}$ and sSFR. In the right panel we plot $M_{\rm dust}/M_{\star}$ ratio vs. sSFR. Here we see an increasing trend, where $M_{\rm dust}/M_{\star}$ ratio increases as sSFR increases. Interestingly, the less massive galaxies seem to follow a parallel path to the more massive ones. In both cases, the LAHz candidates fall in the right side, showing that they have higher sSFR than more massive and dusty galaxies. 

\smallskip

\section{Summary and Conclusions}\label{Sec:Sumary}

\begin{figure*}[hbt] 
\begin{minipage}[l]{1.0\linewidth}
\includegraphics [width=\linewidth]{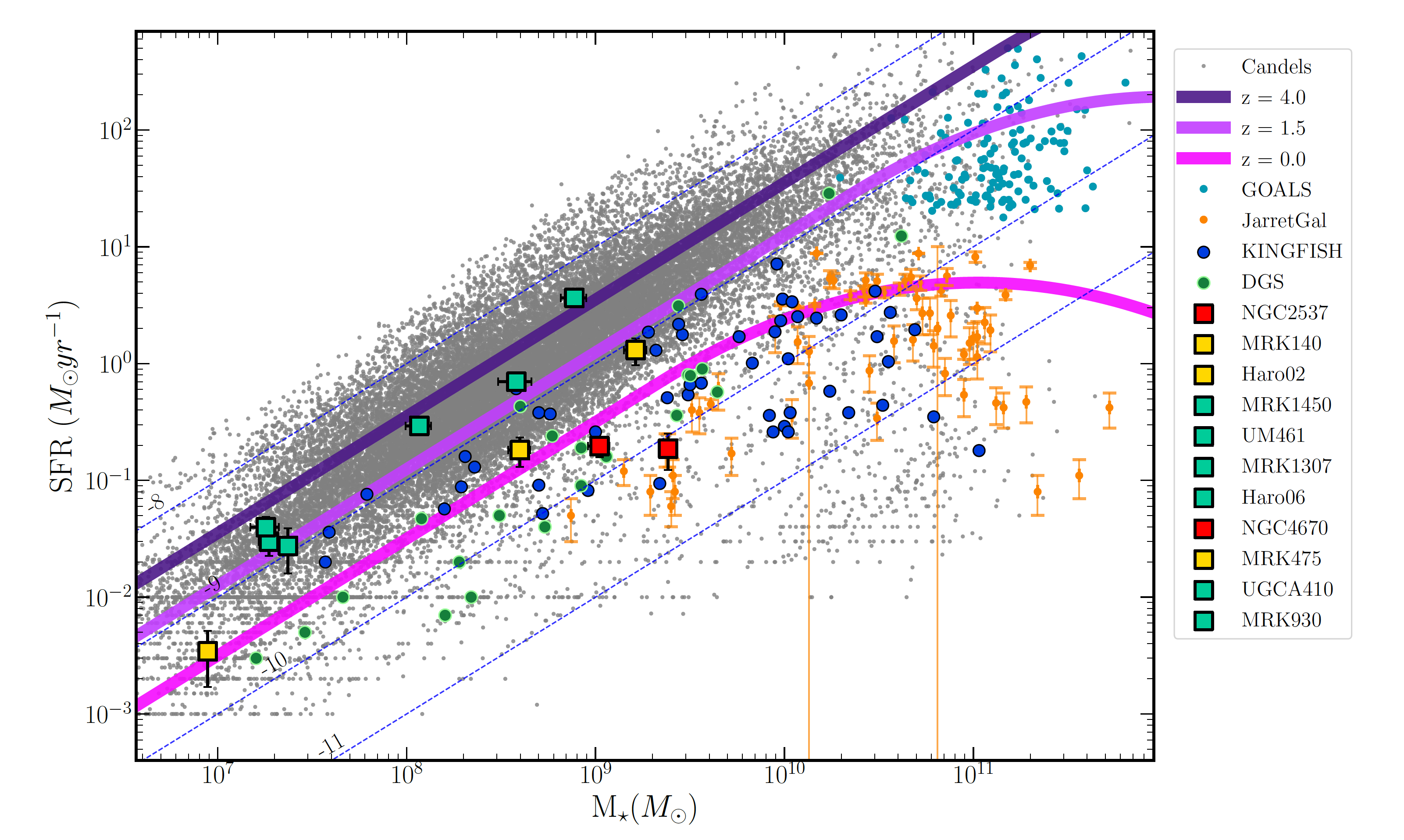}
\caption{Comparison star formation rate (SFR) vs stellar mass ($M_{\star}$), the main sequence (MS) is show in the tick solid line in light-purple for a z=0, purple for z=1.5, and in dark purple for $z=4$. As a comparison, we show the CANDELS galaxies in gray, the {\tt KINGFISH} galaxies in orange, the DGS in cyan, the GOALS in green and the \cite{Jarrett2019} Galaxies in light-green. 
The sample of LAHz (this study), are shown as black squares filled in green whenever they follow the MS for any $z\gtrsim1.5$, in yellow the MS for any $z\gtrsim0$ and filled in red when they follow below the MS for $z=0$. }\label{SFR}   
\end{minipage}
\end{figure*}

The eleven galaxies designated as potential local analogs of high redshift galaxies (LAHz) were initially selected based on the frequency with which they provide the best fits to the SED
of high-$z$ galaxies. They were drawn from the \cite{Brown2014} sample of local galaxy templates, which, although extensive in terms of number of galaxies, only provide a limited set
of dwarf irregular galaxies. It is highly likely that there are other local galaxies with similar properties that could fit into the role as LAHz candidates; they just were not part of the Brown
sample.

\smallskip

As a group, these eleven LAHz galaxies are similar to galaxies in the more diverse Dwarf Galaxy Survey (\citealt{Madden2013}). The LAHz galaxies tend to have higher sSFRs,
larger $M_{\rm dust}/M_{\star}$ ratios, and characterized by warmer dust temperatures, than the average for the DGS sample. Despite the small sample size of LAHz candidates, we found
that they have different properties. The SFHs for six of the eleven LAHz galaxies are consistent with zero star formation activity at times $\gtrsim$1\,Gyr, making them truly young galaxies.
This sub-set of LAHz consists of Mrk\,1450, UM\,461, Mrk\,1307, Mrk\,475, UGCA\,410, and Mrk\,930, and we will refer to it as the `gold' sample. Two of the LAHz galaxies, NGC\,2537 and 
NGC\,4670, have SFHs showing significant star formation activity over a Hubble time. These two galaxies, despite their appearance as Blue Compact Dwarf Galaxies, are therefore 
dominated by an old stellar population, and hence, cannot be representative for high-$z$ galaxies. For the benefit of this discussion, we will refer to these two galaxies as the `bronze'
sample. The remaining three LAHz galaxies; Haro\,02, Haro\,06 and Mrk\,140, have SFHs that cannot rule out star formation activity on longer time scales $\gtrsim$5\,Gyr. Hence,
they fall in between the `gold' and 'bronze' samples, and we will designate them as the `silver' sample.

\smallskip

While we are dealing with small number statistics, there are several notable differences between the `gold' and `bronze' galaxies; the average sSFR of the `gold' sample is $\sim$15
times higher than the average of the two `bronze' galaxies. The average temperature of the cold dust component of the `gold' sample is 48\,K, while both the `silver' and `bronze'
galaxies have an average $T_{\rm cold}$ of 35\,K.
These differences could be due to the stochastic nature of star formation activity in these small galaxies. However, it is also accompanied by a systematic difference in gas-phase metallicity; the `gold' sample contains all the most
metal-poor systems, 0.19\,Z$_{\odot}$, while the `silver' and `bronze' samples have an average metallicity of 0.40\,Z$_{\odot}$. The interstellar gas fractions are 0.68, 0.46 and
0.25 for the `gold', `silver' and `bronze' samples, respectively.
Another parameter setting the `gold' galaxies apart is the flux ratio $S_{24}/S_{160}$, describing the strength of the mid-IR relative to the FIR. These ratios are listed in 
Table~\ref{DustTemperatures}. The `gold' sample have an average ratio $\sim$0.4, compared with 0.07 for NGC\,2537 and NGC\,4670. 

\smallskip

In Figure~\ref{SFR} we plot the SFR vs stellar mass for our LAHz galaxies, and several other surveys; including the DGS and {\tt KINGFISH} samples; the {\tt GOALS} sample of ultra-luminous infrared galaxies (\citealt{Armus2009PASP..121..559A}), and the \cite{Jarrett2019} sample of the 100 largest nearby galaxies. In addition plot high redshift galaxies with $1 < z < 4.5$ from the CANDELS survey (\citealt{Santini2015}). Diagonal dashed lines show different sSFR values, and the thick diagonal lines show the main sequence for redshifts $z$=0, 1.5 and 4 (\citealt{Schreiber2015}). 
Our LAHz galaxies are shown as squares, with colors corresponding to whether they belong in the `'gold' (green), `silver' (red), or `bronze' (orange) samples. Most of the
DGS, {\tt KINGFISH}, and \cite{Jarrett2019} galaxies are centered around the main sequence (MS) for $z$=0, including the turn-off at high stellar masses. The {\tt GOALS}
galaxies have, not surprisingly, a higher sSFR. The `gold' LAHz galaxies have high sSFR, and are located above the $z$=0 MS. The two `bronze' LAHz (NGC\,2537 and
NGC\,4670), are located below the $z$=0 MS, and the three `silver' LAHz are on the $z$=0 MS.  Hence, the six LAHz in the `gold' sample can be viewed either as $z$=0
starburst systems, or as MS galaxies at $z\sim$1.5-4.

\smallskip

Based on these findings, we propose that the six galaxies in the `gold' sample are good candidates for follow-up studies, especially determining their gas dynamics.
If they truly are young systems, their age puts their formation epoch at $z$$\sim$0.08, or a luminosity distance of only $\sim$350\.Mpc. This short time scale means
that gravitational interaction and merging are not likely to be driving the star formation activity.

\smallskip

\begin{acknowledgments}
Based on observations made with the NASA/DLR Stratospheric Observatory for Infrared Astronomy (SOFIA). SOFIA is jointly operated by the Universities Space Research
Association, Inc. (USRA), under NASA contract NNA17BF53C, and the Deutsches SOFIA Institut (DSI) under DLR contract 50 OK 0901 to the University of Stuttgart. Financial
support for this work was provided by NASA through award \#06-0222 issued by USRA. We extend our thanks to the SOFIA science support team, for all the help in the data acquisition and the data analysis process. 

\end{acknowledgments}


\input{LocalAnalogsPaper_Bib_AAStex.tex}

\end{document}

%% file: Table1_SampleDescription.tex

\begin{table*} [!htb]
\begin{center}
\caption{Properties of the local analogs} \label{SampleDescription}
\begin{tabular}{l c c c c c c c c c c c }
\hline   
\hline     
\sc{No.} & \sc{Galaxy}  &  RA & Dec  & D   & $\log{M_*}^{(a)}$   & $\log{M_{HI}}^{(b)}$ & f$_{\mathrm{gas}}^{(c)}$ & Metallicity$^{(d)}$ & Alternative &      & \\
&\sc{Name}  & \multicolumn{2}{c}{ J2000.0}  & Mpc & M$_{\odot}$  & M$_{\odot}$     & M$_{\odot}$       &   $12+\log{(O/H)}$ &    Name     & \\
\hline 
    1     &  NGC 2537         &      08:13:14.4   &   +45:59:13     &      8.6           &  9.02 & 8.43     &  0.17   & 8.19 &   Arp~6; `Bear Paw'     \\  
      2     &  Mrk 140          &      10:16:28.3   &   +45:19:18     &     27.6           &  8.60 & 8.90     &  0.56   & 8.30 &    Mrk~140        \\  
      3     &  Haro 02          &      10:32:31.9   &   +54:24:02     &     23.7           &  9.21 & 8.03     &  0.02   & 8.45 &   Mrk~033        \\  
      4     &  Mrk 1450         &      11:38:35.6   &   +57:52:27     &     14.7           &  7.27 & 7.35     &  0.54   & 7.96 &                      \\  
      5     &  UM 461           &      11:51:33.1   &   -02:22:22     &     20.7           &  7.37 & 8.47     &  0.96   & 7.78 &                      \\  
      6     &  Mrk 1307         &      11:52:37.4   &   -02:28:09     &     21.0           &  8.07 & 8.73     &  0.81   & 7.96 &    UM~462         \\  
      7     &  Haro 06          &      12:15:18.4   &   +05:45:39     &     35.1           &  8.58 & 8.83     &  0.69   & 8.18 &                     \\  
      8     &  NGC 4670      &      12:45:17.1   &   +27:07:31     &     20.0           &  9.38 & 9.02     &  0.37   & 8.30 &   Arp~163         \\  
      9     &  Mrk 475          &      14:39:05.5   &   +36:48:21     &     10.9           &  6.95 & 6.62     &  0.56   & 7.93 &                     \\  
     10     &  UGCA 410     &      15:37:04.2   &   +55:15:48     &     10.5           &  7.26 & 7.58     &  0.57   & 8.10 &    Mrk~487     \\  
     11     &  Mrk 930          &      23:31:58.6   &   +28:56:50     &     77.5           &  8.89 & 9.51     &  0.74   & 8.08 &                  \\  
\hline    
\hline
\end{tabular}
\end{center}
{(a)\ {\em Stellar mass} ${M_*}$ This work.
 (b) \ {\em HI} data from Paturel et al. (2003), except for Mrk~1450 (\cite{vanDriel2016A}). 
 (c) \ {\em Gas fraction} is defined as $f_g = M_g/(M_g + M_*)$.
 (d) \ {\em Metal Abundances}  UM~461 and UM~462 (\textcite{Campos-Aguilar1993}), Haro 02 (\textcite{Davidge1989}).
}
\end{table*}

%% file: Table2_DataAvailable.tex
\begin{tabular}{l l c c c c c c c c c c c c c c c }
\hline\hline   
%
%
\multicolumn{1}{c}{ }  &
\multicolumn{1}{c}{ \sc{Galaxy}}  &
\multicolumn{14}{c}{\sc{FIR Photometric Data}  $(\mu m)$ } \\
%
%
\multicolumn{1}{c}{  \sc{No}      }&
\multicolumn{1}{c}{  \sc{Name} }&
\multicolumn{1}{c}{  12-22  }&
\multicolumn{1}{c}{ 24-25 }&
\multicolumn{1}{c}{  55 }&
\multicolumn{1}{c}{  60  } &
\multicolumn{1}{c}{  65  }&
\multicolumn{1}{c}{ 70 }&
\multicolumn{1}{c}{ 89-90 }&
\multicolumn{1}{c}{ 100 }& 
\multicolumn{1}{c}{ 140-155 }&
\multicolumn{1}{c}{ 160 }& 
\multicolumn{1}{c}{ 214 }& 
\multicolumn{1}{c}{ 250 }& 
\multicolumn{1}{c}{ 350 }& 
\multicolumn{1}{c}{ 500  } \\
%
%
\hline
%
%
              1   & NGC 2537  &  W                &   M    &  S    & I     &  -     &   M    &  S      &   I      &  S  & M   &  -   &  -  &  -    &  -   \\
              2   & Mrk 140     &  W                 &   M    &  -     & I     &  -     &   H    &  -        &   H    &  -  & H   &  -   &  H  &  H   &  H    \\
              3   & Haro 02      &  W                &  M-I   &  -     & I     &  -     &   H    &  -        &   H    & -   & H    &  -   &  H  &  H   &  H  \\
              4   & Mrk 1450    &  W                &   M    &  S    & -    &  -     &   H     &  S      &   H    &  S  & H   &  S   &  H  &  H   &  H   \\ 
              5   & UM 461      & IRS16 - W    &   M    &  -     & I     &  -     &   H    &  -        &   H    &  -  & H   &  -   &  H  &  H   &  H    \\
              6   & Mrk 1307    &  W                &   M    &  S    & I     &  -    &   M    &  S       &   I     &  S  & M   &  -   &  -  &  -    &  -   \\
              7   & Haro 06      &  W                &   M    &  S    & I     &  -    &   H     &  S       &   H   &  S  & H   &  S   &  H  &  H   &  H    \\
              8   & NGC 4670  &  W                &   M    &  S    & I     &  -    &   M    &  S       &   I     &  S  &  M    &  S   &  -  &  -    &  -   \\ 
              9   & Mrk 475      & IRS16 - W    &   M    &  -     &  -    &  -    &   M    &  -        &   -     & -   & M    &  -   &  -  &  -    &  - \\  
              10 & UGCA 410      & IRS16          &   M-I  &  -     &  I    &  A    &   -      &  A       &   -     &  -  & A    & -   &  -  &  -    &  - \\
              11 & Mrk 930      &  W                &   M    & -      &  -    &  -     &   H    &  -        &   H    &  -  & H   &  -   &  H  &  H   &  H    \\ 
\hline\hline 
%
%
\end{tabular}

%% file: Table3_PhotSOFIA.tex
\begin{tabular}{l c c c c c c c c c }  
\hline\hline 
\multicolumn{1}{c}{\sc{Galaxy}} & 
\multicolumn{1}{c}{ HAWC+A } & 
\multicolumn{1}{c}{ MIPS } & 
\multicolumn{1}{c}{ HAWC+C } & 
\multicolumn{1}{c}{ PACS }  &
\multicolumn{1}{c}{ HAWC+D } & 
\multicolumn{1}{c}{ HAWC+E } & 
\multicolumn{1}{c}{ SPIRE }  \\ 
%
%
\multicolumn{1}{c}{\sc {Name}} & 
\multicolumn{1}{c}{ $55\,\mu m$ } & 
\multicolumn{1}{c}{ $60\,\mu m$ } & 
\multicolumn{1}{c}{ $89\,\mu m$ } & 
\multicolumn{1}{c}{ $100\,\mu m$ } & 
\multicolumn{1}{c}{ $155\,\mu m$ } & 
\multicolumn{1}{c}{ $214\,\mu m$ } & 
\multicolumn{1}{c}{ $250\,\mu m$ }  \\ 
%
%
\hline 
\vspace{-0.05in} \\ 
NGC 2537 & $  1302.4\pm    255.0$& $  4040.0\pm    286.4$& $  4893.3\pm    914.1$  &   -  & $  6301.9\pm   1085.0$  &   -    &   -    \\ 
MRK 140    &   -    &   -    &   -  & $   563.2\pm     44.8$  &   -    &   -  & $   172.0\pm     26.3$  \\ 
Haro 02    &   -    &   -    &   -  & $  5051.3\pm    254.6$  &   -    &   -  & $   911.9\pm    137.2$  \\ 
MRK 1450 & $   274.8\pm     41.6$  &   -  & $   317.0\pm     47.9$& $   272.0\pm     21.8$& $   168.2\pm     27.4$& $    52.5\pm     17.3$& $    48.7\pm      8.4$  \\ 
UM 461     &   -    &   -    &   -  & $   132.0\pm     12.7$  &   -    &   -  & $    23.1\pm      7.1$  \\ 
MRK 1307 & $   724.7\pm    119.9$& $  1000.0\pm     71.6$& $  1277.6\pm    198.6$  &   -  & $   616.5\pm    115.4$  &   -    &   -    \\ 
Haro 06  & $   646.3\pm     97.0$  &   -  & $   849.4\pm    127.4$& $   742.3\pm     39.2$& $   754.3\pm    115.9$& $   309.8\pm     48.7$& $   197.3\pm     30.0$  \\ 
NGC 4670 & $  1428.2\pm    246.1$& $  3470.0\pm    245.7$& $  3505.9\pm    530.7$  &   -  & $  3531.9\pm    713.4$& $  2486.7\pm    428.9$  &   -    \\ 
MRK 475    &   -  & $   110.0\pm     16.0$  &   -    &   -    &   -    &   -    &   -    \\ 
UGCA 410   &   -    &   -    &   -    &   -    &   -    &   -    &   -    \\ 
MRK 930    &   -    &   -    &   -  & $  1140.0\pm     91.9$  &   -    &   -  & $   210.4\pm     31.9$  \\ 
\hline\hline 
\end{tabular} 

%% file: Table5_BayesianParameters.tex
%
%
\begin{tabular}{l l c c c c c} 
\hline \hline   
%
%
 \multicolumn{1}{l}{\sc{No}}     & 
 \multicolumn{1}{c}{\sc{Galaxy}}     & 
 \multicolumn{1}{c}{$T_{d\,Warm}$  }     & 
 \multicolumn{1}{c}{$T_{d\,Cold } $  }     & 
 \multicolumn{1}{c}{$S_{24\mu m}/S_{160\mu m}$\tablenotemark{1} }     & 
 \multicolumn{1}{c}{$M_{d}$  }     & 
 \multicolumn{1}{c}{$L_{FIR}$}      \\
%
 \multicolumn{1}{l}{       }     & 
 \multicolumn{1}{c}{\sc{Name}  }     & 
 \multicolumn{1}{c}{K}     & 
 \multicolumn{1}{c}{ K}     & 
 \multicolumn{1}{c}{}     & 
 \multicolumn{1}{c}{$M_{\odot}$}     & 
 \multicolumn{1}{c}{$10^{9}\,L_{\odot}$}       \\ 
\hline\hline   

	    1& NGC 2537	& $   139.0_{-16}^{+70}   $	& $    34.0_{-1.5}^{+0.9}   $   & $ 0.06$    & $ 3.83_{-0.8}^{+0.7} \times10^{ 5}$ 	& $0.638 \pm {0.028} $  \\  
   \vspace{-0.1in} \\ 
	    2& Mrk 140 	& $    79.0_{- 7}^{+ 5}   $	& $    31.0_{-3.6}^{+1.0}   $   & $ 0.10$    & $ 4.05_{-0.5}^{+1.1} \times10^{ 5}$ 	& $0.784 \pm {0.043} $  \\  
   \vspace{-0.1in} \\ 
	    3& Haro 02 	& $   153.0_{-29}^{+ 8}   $	& $    42.5_{-1.6}^{+0.8}   $   & $ 0.29$    & $ 9.53_{-1.0}^{+0.7} \times10^{ 5}$ 	& $6.222 \pm {0.172} $  \\  
   \vspace{-0.1in} \\ 
	    4& Mrk 1450	& $   100.0_{- 9}^{+37}   $	& $    44.0_{-4.5}^{+5.3}   $   & $ 0.41$    & $ 1.36_{-0.3}^{+0.2} \times10^{ 4}$ 	& $0.150 \pm {0.008} $  \\  
   \vspace{-0.1in} \\ 
	    5& UM 461  	& $   148.0_{-15}^{+ 8}   $	& $    46.0_{-3.8}^{+2.5}   $   & $ 0.42$    & $ 1.55_{-0.3}^{+0.4} \times10^{ 4}$ 	& $0.161 \pm {0.009} $  \\  
   \vspace{-0.1in} \\ 
	    6& Mrk 1307	& $   149.0_{-40}^{+ 1}   $	& $    46.0_{-4.2}^{+0.9}   $   & $ 0.23$    & $ 1.12_{-0.2}^{+0.4} \times10^{ 5}$ 	& $0.913 \pm {0.038} $  \\  
   \vspace{-0.1in} \\ 
	    7& Haro 06 	& $    79.0_{- 2}^{+10}   $	& $    32.0_{-1.6}^{+3.1}   $   & $ 0.14$    & $ 6.67_{-1.4}^{+0.6} \times10^{ 5}$ 	& $1.809 \pm {0.092} $  \\  
   \vspace{-0.1in} \\ 
	    8& NGC 4670	& $   162.0_{-51}^{+13}   $	& $    36.0_{-2.5}^{+0.4}   $   & $ 0.08$    & $ 1.19_{-0.2}^{+0.3} \times10^{ 6}$ 	& $2.665 \pm {0.109} $  \\  
   \vspace{-0.1in} \\ 
	    9& Mrk 475 	& $   119.0_{-11}^{+ 8}   $	& $    51.0_{-8.6}^{+4.3}   $   & $ 0.24$    & $ 1.96_{-0.7}^{+1.6} \times10^{ 3}$ 	& $0.024 \pm {0.003} $  \\  
   \vspace{-0.1in} \\ 
	   10& UGCA 410	& $   146.0_{-10}^{+29}   $	& $    58.0_{-4.7}^{+5.8}   $   & $ 0.71$    & $ 3.71_{-1.0}^{+0.9} \times10^{ 3}$ 	& $0.103 \pm {0.006} $  \\  
   \vspace{-0.1in} \\ 
	   11& Mrk 930 	& $    94.0_{-14}^{+ 6}   $	& $    40.0_{-8.3}^{+1.4}   $   & $ 0.29$    & $ 2.18_{-0.3}^{+0.7} \times10^{ 6}$ 	& $16.225 \pm {1.313} $  \\  
   \vspace{-0.1in} \\ 
\hline \hline 	 
%
%
\end{tabular}
\tablenotetext{1}{The flux ratio $S_{24\,\mu m}/S_{160\,\mu m}$ is derived from the modified black-body best fit model.}

%% file: Table6_SFR_SFH.tex
%
%
\begin{tabular}{l l r r r r r} 
\hline \hline  
 \multicolumn{1}{l}{No.}   & 
 \multicolumn{1}{c}{\sc{Galaxy}}   & 
 \multicolumn{1}{c}{$M_{dust}$    }   & 
 \multicolumn{1}{c}{$L_{FIR}$  }   & 
 \multicolumn{1}{c}{$M_{*}$    }   & 
 \multicolumn{1}{c}{SFR\tablenotemark{1}  }   & 
 \multicolumn{1}{c}{sSFR           }     \\ 
 \multicolumn{1}{l}{  }   & 
 \multicolumn{1}{c}{\sc{Name} }   & 
 \multicolumn{1}{c}{$10^{4}\,\rm M_{\odot}$   }   & 
 \multicolumn{1}{c}{$10^{9}\,\rm L_{\odot}$}   & 
 \multicolumn{1}{c}{$10^{8}\,\rm M_{\odot}$}   & 
 \multicolumn{1}{c}{$\rm M_{\odot}\,yr^{-1}$}   & 
 \multicolumn{1}{c}{$\rm Gyr^{-1}$}     \\ 
\hline 
\vspace{-0.1in} \\  
    1& NGC 2537       &    $ 9.5\pm\,3.0       $ &  $    0.77_{-0.02}^{+0.02}  $ &  $ 10.5_{-1.4}^{+1.3} $ &  $    0.20_{-0.04}^{+0.04}$  &   $ 0.19\pm\,0.04$   \\  
   \vspace{-0.1in} \\ 
    2& Mrk 140          &    $ 25.7\pm\,7.2      $ &  $    0.94_{-0.03}^{+0.03}  $ &  $ 3.9_{-0.5}^{+0.5} $ &  $    0.18_{-0.05}^{+0.05}$  &   $ 0.46\pm\,0.14$       \\  
   \vspace{-0.1in} \\ 
    3& Haro 02          &    $ 21.5\pm\,4.8      $ &  $    8.03_{-0.20}^{+0.21}  $ &  $ 16.3_{-2.2}^{+2.2} $ &  $    1.3_{-0.3}^{+0.4}$  &   $ 0.80\pm\,0.23$   \\  
   \vspace{-0.1in} \\ 
    4& Mrk 1450        &    $ 1.9\pm\,0.2        $ &  $    0.19_{-0.01}^{+0.01}  $ &  $ 0.19_{-0.02}^{+0.03} $ &  $    0.03_{-0.01}^{+0.01}$  &   $ 1.60\pm\,0.44$   \\  
   \vspace{-0.1in} \\ 
    5& UM 461          &    $ 1.3\pm\,0.3        $ &  $    0.21_{-0.01}^{+0.01}  $ &  $ 0.21_{-0.02}^{+0.03} $ &  $    0.03_{-0.01}^{+0.01}$  &   $ 1.17\pm\,0.50$   \\  
   \vspace{-0.1in} \\ 
    6& Mrk 1307        &    $ 11.0\pm\,2.9      $ &  $    1.10_{-0.04}^{+0.04}  $ &  $ 1.2_{-0.2}^{+0.2} $ &  $    0.29_{-0.04}^{+0.05}$  &   $ 2.51\pm\,0.55$   \\  
   \vspace{-0.1in} \\ 
    7& Haro 06           &    $ 81.8\pm\,5.8     $ &  $    2.18_{-0.06}^{+0.06}  $ &  $ 3.8_{-0.7}^{+0.8} $ &  $    0.70_{-0.06}^{+0.06}$  &   $ 1.83\pm\,0.40$   \\  
   \vspace{-0.1in} \\ 
    8& NGC 4670       &    $ 207.0\pm\,66.0 $ &  $    3.30_{-0.1}^{+0.1}  $ &  $ 24.2_{-2.1}^{+1.9} $ &  $    0.19_{-0.06}^{+0.07}$  &   $ 0.08\pm\,0.03$   \\  
   \vspace{-0.1in} \\ 
    9& Mrk 475           &    $ 0.17\pm\,0.04  $ &  $    0.028_{-0.002}^{+0.002}  $ &  $ 0.09_{-0.01}^{+0.01} $ &  $    0.003_{-0.001}^{+0.002}$  &   $ 0.39\pm\,0.20$   \\  
   \vspace{-0.1in} \\ 
   10& UGCA 410      &    $ 2.6\pm\,1.2      $ &  $    0.15_{-0.01}^{+0.01}  $ &  $ 0.18_{-0.03}^{+0.04} $ &  $    0.04_{-0.01}^{+0.01}$  &   $ 2.21\pm\,0.64$   \\  
   \vspace{-0.1in} \\ 
   11& Mrk 930          &    $ 68.7\pm\,19.5  $ &  $   19.1_{-0.7}^{+0.7}  $ &  $ 7.7_{-1.1}^{+1.3} $ &  $    3.6_{-0.4}^{+0.4}$  &   $ 4.71\pm\,0.87$   \\  
   \vspace{-0.1in} \\ 
\hline \hline 
%
%
\end{tabular}

%% file: LocalAnalogsPaper_Bib_AAStex.tex
%

\bibliography{sample63}{}
\bibliographystyle{aasjournal}

%